\documentclass[twocolumn]{aastex631_arxiv}

\usepackage{booktabs}
\usepackage{multirow}
\usepackage{xcolor}
\usepackage{soul}
\usepackage{chngcntr}
\usepackage{placeins}
\usepackage{bm}

\received{August 12, 2022}
\revised{September 16, 2022}
\accepted{September 20, 2022}

\submitjournal{the Planetary Science Journal}

\shorttitle{Predicting asteroid properties from impact}
\shortauthors{Kumamoto et al.}
\graphicspath{{./}{figs/}}

\begin{document}

\title{Predicting asteroid material properties from a DART-like kinetic impact}


\correspondingauthor{Kathryn M. Kumamoto}
\email{kumamoto3@llnl.gov}

\author[0000-0002-0400-6333]{Kathryn M. Kumamoto}
\affiliation{Lawrence Livermore National Laboratory, 7000 East Avenue, Livermore, CA 94550}

\author[0000-0003-4796-124X]{J.~Michael Owen}
\affiliation{Lawrence Livermore National Laboratory, 7000 East Avenue, Livermore, CA 94550}

\author{Megan Bruck Syal}
\affiliation{Lawrence Livermore National Laboratory, 7000 East Avenue, Livermore, CA 94550}

\author{Jason Pearl}
\affiliation{Lawrence Livermore National Laboratory, 7000 East Avenue, Livermore, CA 94550}

\author{Cody Raskin}
\affiliation{Lawrence Livermore National Laboratory, 7000 East Avenue, Livermore, CA 94550}

\author[0000-0001-6076-5636]{Wendy K. Caldwell}
\affiliation{Los Alamos National Laboratory, P.O. Box 1663, Los Alamos, NM 87545}

\author{Emma Rainey}
\affiliation{Applied Physics Laboratory, Johns Hopkins University, 11100 Johns Hopkins Road, Laurel, MD 20723}

\author[0000-0002-7602-9120]{Angela Stickle}
\affiliation{Applied Physics Laboratory, Johns Hopkins University, 11100 Johns Hopkins Road, Laurel, MD 20723}

\author{R.~Terik Daly}
\affiliation{Applied Physics Laboratory, Johns Hopkins University, 11100 Johns Hopkins Road, Laurel, MD 20723}

\author[0000-0002-3578-7750]{Olivier Barnouin}
\affiliation{Applied Physics Laboratory, Johns Hopkins University, 11100 Johns Hopkins Road, Laurel, MD 20723}

\begin{abstract}

NASA's Double Asteroid Redirection Test (DART) mission is the first full-scale test of the kinetic impactor method for asteroid deflection, in which a spacecraft intentionally impacts an asteroid to change its trajectory. DART represents an important first step for planetary defense technology demonstration, providing a realistic assessment of the effectiveness of the kinetic impact approach on a near-Earth asteroid. The momentum imparted to the asteroid is transferred from the impacting spacecraft and enhanced by the momentum of material ejected from the impact site. However, the magnitude of the ejecta contribution is dependent on the material properties of the target. These properties, such as strength and shear modulus, are unknown for the DART target asteroid, Dimorphos, as well as most asteroids since such properties are difficult to characterize remotely.

This study examines how hydrocode simulations can be used to estimate material properties from information available post-impact, specifically the asteroid size and shape, the velocity and properties of the impacting spacecraft, and the final velocity change imparted to the asteroid. Across $>$300 three-dimensional simulations varying seven material parameters describing the asteroid, we found many combinations of properties could reproduce a particular asteroid velocity. Additional observations, such as asteroid mass or crater size, are required to further constrain properties like asteroid strength or outcomes like the momentum enhancement provided by impact ejecta. Our results demonstrate the vital importance of having as much knowledge as possible prior to an impact mission, with key material parameters being the asteroid's mass, porosity, strength, and elastic properties.

\end{abstract}



\section{Introduction}

Planetary defense preparedness depends upon the ability of space-faring nations to deploy mature mitigation technologies in a timely manner in the event of an Earth-impact emergency. As outlined in the United States' National Near Earth Object Preparedness Strategy and Action Plan \citep{OSTP2018}, further development of deflection and disruption technologies is required before an imminent threat arises. Kinetic impact is a relatively mature asteroid deflection technology, owing to its simplicity: a spacecraft impacting at many kilometers per second can deliver a significant momentum impulse to an asteroid, through both its own mass and velocity and the additional ``boost" of momentum from escaping crater ejecta. As a result of launch vehicle mass limitations, kinetic impact is typically not a viable strategy for shorter warning time scenarios or asteroids that exceed a few hundred meters in diameter \citep{dearborn2015,dearborn2020}. However, the range of scenarios over which kinetic impact can be effective will partly depend upon the momentum multiplier provided by escaping crater ejecta. For this reason, in addition to demonstrating targeting capabilities and operational readiness, it is desirable to conduct kinetic impact experiments on real asteroids, in which the imparted change in momentum can be well quantified. 

Asteroid characteristics that may contribute to uncertainty in an impulsive deflection response include mass, strength, porosity, shape, internal structure, spin state, and equation of state \citep[e.g.,][]{asphaug1998,brucksyal2016,feldhacker2017,holsapple2012,jutzi2014,raducan2019,raducan2020}. Currently, a lack of direct data for asteroid material properties, combined with the likely lack of target-specific data on future threats and the observable diversity across the population of Near-Earth Objects (NEOs), contributes to uncertainty in the momentum multiplication from future kinetic impactor missions. While the constitutive properties of meteorites can be carefully studied in Earth-based laboratories \citep[e.g.,][]{cottofigueroa2016,kimberley2011,moyanocambero2017,flynn2018}, these samples are biased toward materials that survive atmospheric entry and impact. Additionally, the rubble-pile structures of many NEOs include significant macroporosity, producing bulk geotechnical properties that are likely distinct from the properties of meteoritic hand samples. Both integrated experiments, such as a kinetic impact test with a measurable deflection velocity, and focused science experiments, such as in-situ or sample-return characterization, can provide critical ground-truth information for asteroid material properties.

The Asteroid Impact and Deflection Assessment (AIDA) is a collaboration between NASA and ESA to study the effects of impacting an asteroid, increasing Earth's preparedness for potentially hazardous asteroids \citep{cheng2015,cheng2016,cheng2018}. It is composed of NASA's Double Asteroid Redirection Test (DART) mission and ESA's Hera mission, as well as the Light Italian CubeSat for Imaging of Asteroids \citep [LICIACube; ][]{dotto2021} carried by DART. In the AIDA collaboration, the DART spacecraft acts as a kinetic impactor, LICIACube follows a few minutes behind to image the ejecta cone soon after impact, and Hera characterizes the asteroid a few years later. DART will impact Dimorphos, the secondary asteroid of the binary system 65803~Didymos. One of the earliest observable consequences of the DART impact will be the change in the orbital period of Dimorphos around the primary asteroid Didymos, which will be measurable from Earth-based telescopes. This change in asteroid velocity is one of the key outcomes of a planetary defense deflection mission, for which the goal would be to nudge the asteroid onto a trajectory that would not impact Earth. DART represents an important planetary defense milestone, as it will carry out Earth's first asteroid deflection test. Interpretation of the DART experiment's results, through modeling and simulation comparison, is essential for estimation of crater ejecta momentum and for the related question of asteroid initial conditions at the impact site. 

In support of the DART/AIDA mission, this study presents the results of an inverse test designed to explore the predictive power of the limited information likely available in the weeks following the DART impact. In this test, the Applied Physics Laboratory at Johns Hopkins University (JHUAPL) and Los Alamos National Laboratory (LANL) ran full impact simulations to act as the ``truth simulation." The asteroid velocity change resulting from this impact simulation was provided to a team at Lawrence Livermore National Laboratory, along with information regarding the impacting spacecraft, a shape model for the target asteroid, and the impact location on that target, but without any information regarding the target asteroid's material properties. Using this limited dataset, we ran impact simulations covering a wide range of material properties to find combinations of parameters that would satisfy the provided velocity change. The inverse test can be viewed as an intensive ``dress rehearsal" for impact modelers to prepare for the actual DART experiment, with the truth simulation taking the place of the actual DART impact.

\section{Information from the truth simulation}

\begin{figure*}
\gridline{\fig{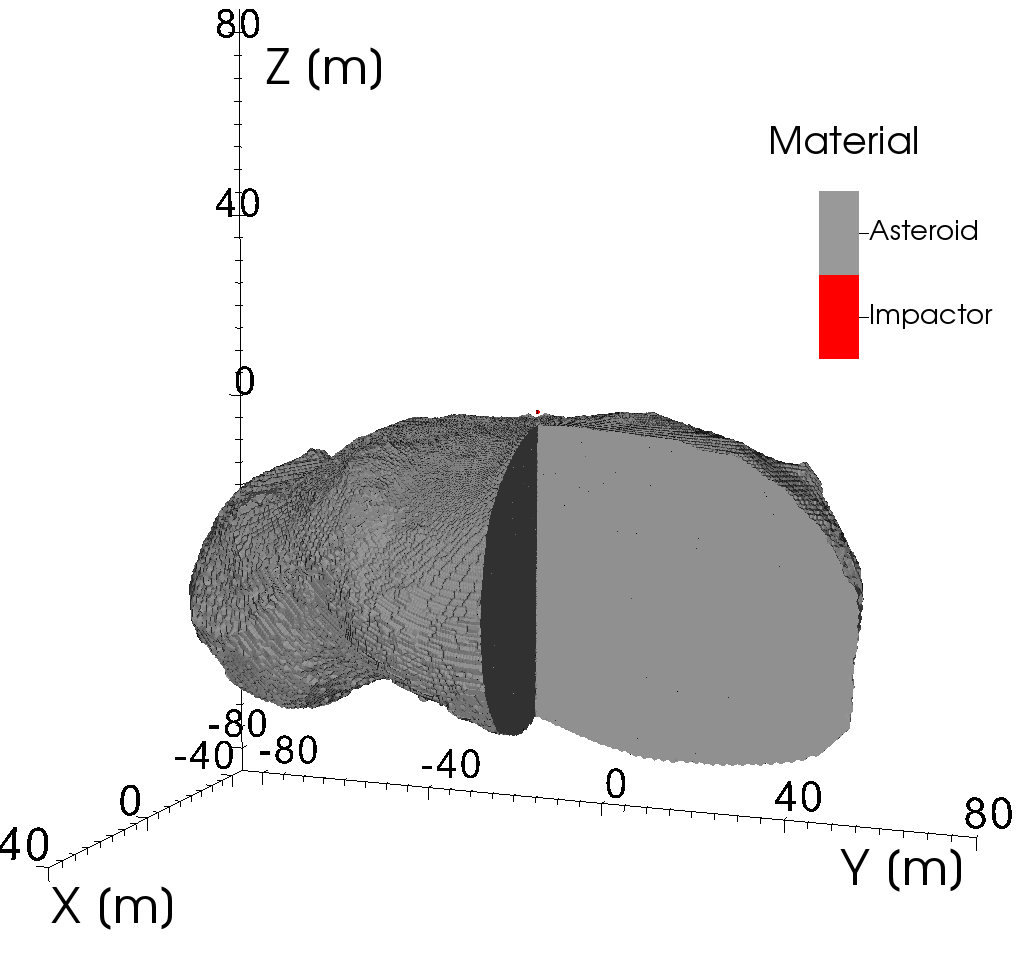}{0.45\textwidth}{(a)}
		  \fig{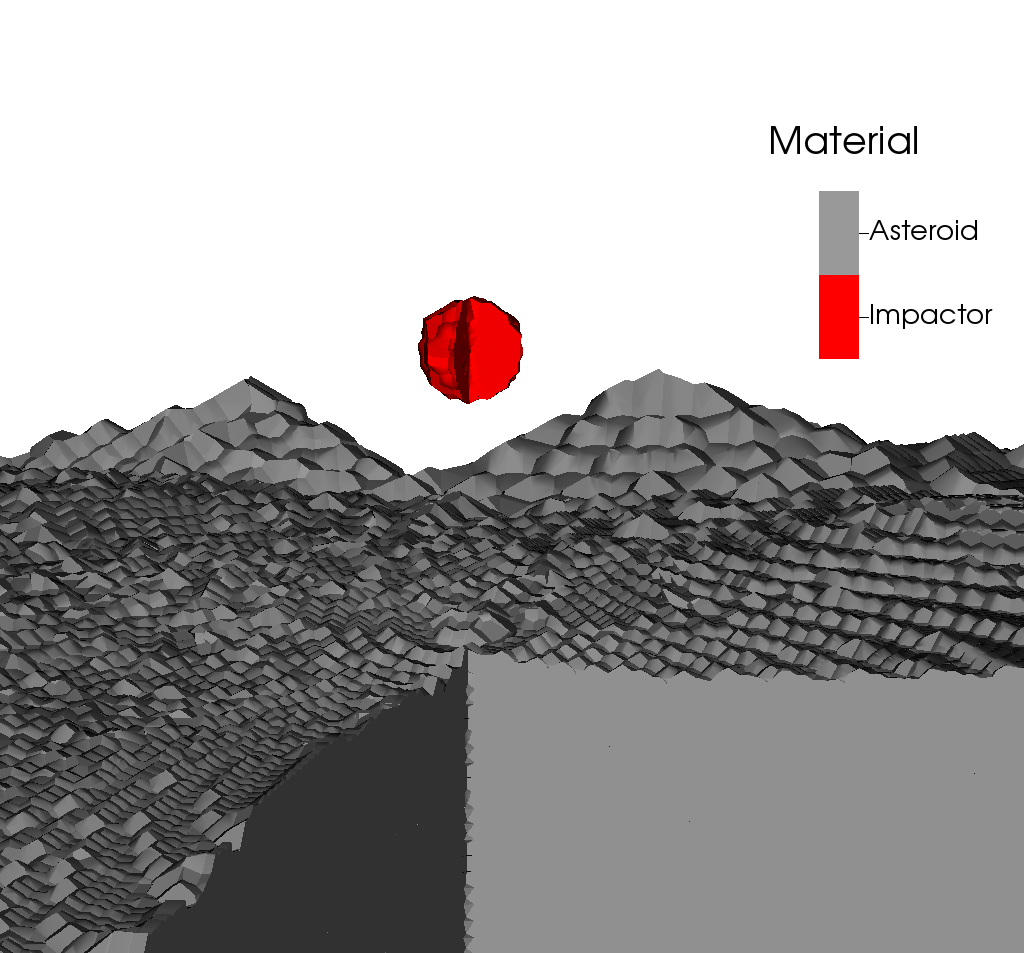}{0.45\textwidth}{(b)}
          }
\gridline{\fig{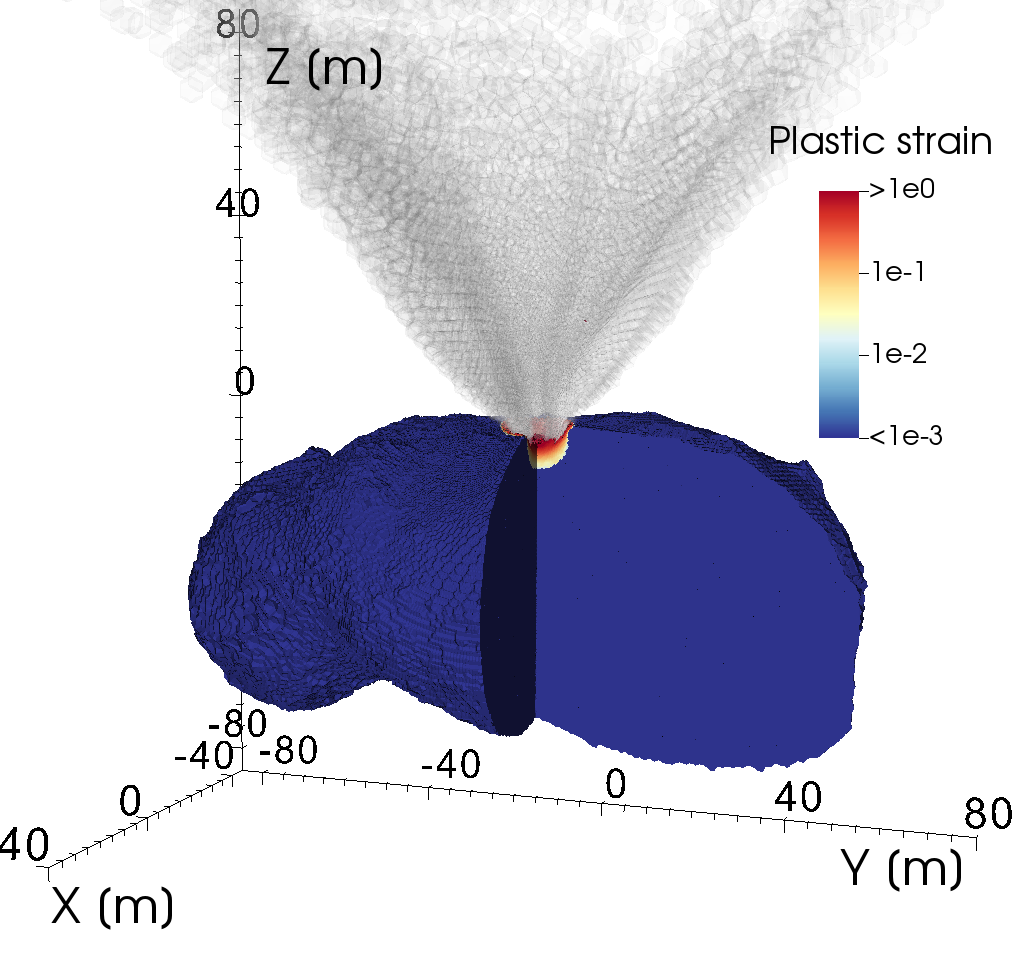}{0.45\textwidth}{(c)}
          \fig{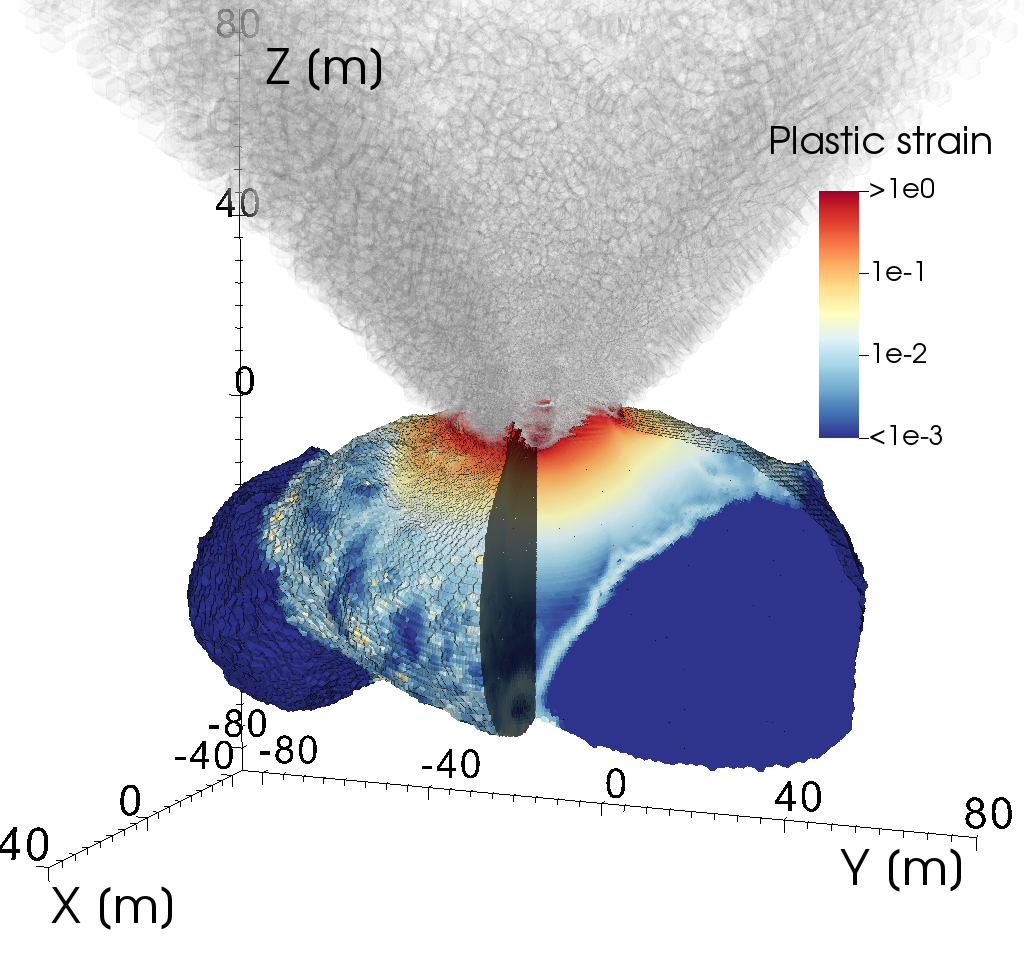}{0.45\textwidth}{(d)}
          }
\caption{Example visualizations of the Spheral simulations. One quadrant is clipped out of the simulation to show the interior structure of the asteroid. (a) The setup for our simulations. The asteroid body is in grey, and the impactor is in red. (b) Same as in (a) but zoomed in on the impactor. (c) The results from simulation 11.09. The asteroid is colored by plastic strain, and low-density ejecta is translucent grey. This simulation had a relatively strong and porous asteroid body. (d) The results from simulation 22.05. Colors are the same as in (c). This simulation had a relatively weak asteroid body with low porosity.}
\label{fig_visit}
\end{figure*}

For this inverse test, the target asteroid was modeled using a shape model of the asteroid 25143~Itokawa \citep{fujiwara2006}, shrunken such that the long axis of the asteroid was $\sim$130~m, giving a total volume of $\sim$280,000~m$^3$ (Figure \ref{fig_visit}a). This shape model was constructed in \citet{daly2022} from images simulated to match those received from the DART mission, including both the camera onboard DART \citep[the Didymos Reconnaissance and Asteroid Camera for Optical navigation, DRACO; ][]{fletcher2018} and a camera on LICIACube \citep[the LICIACube Explorer Imaging for Asteroid, LEIA; ][]{dotto2021}. Itokawa was chosen for this inverse test as it is an S-type asteroid like Didymos (and as expected for Dimorphos) and may therefore have similar surface attributes, topography, and characteristics. The aspect ratio of Itokawa, however, is larger than that expected for Dimorphos, resulting in the total volume of the shape model being $\sim$12\% of that predicted for Dimorphos.

The truth simulation of the impact was run using two hydrocodes: CTH \citep{mcglaun1990} at JHUAPL and Free LAGrange \citep[FLAG; ][]{burton1992,burton1994a,burton1994b} at LANL. The material properties of the asteroid are presented in Table \ref{tab_res}, but parameters such as target strength and porosity were not provided to the inverse modelers since the properties of Dimorphos will also be unknown prior to impact. The spacecraft velocity vector, also used for the simulated DRACO images to construct the shape model \citep{daly2022}, was slightly oblique to the surface of the target, with a mean angle of 9.5$^{\circ}$ between the spacecraft velocity and the surface normal at the point of impact. Figures~\ref{fig_visit}a and b demonstrate the model setup, with the asteroid body in grey and the impactor in red. Using a spacecraft with a mass of 612~kg and relative speed of 7.2~km/s, the impact caused the body's orbital velocity to decrease by 1.972~$\pm$~0.07~cm/s in the truth simulation, reducing its orbital period around the primary from 12 to 6 hours. For more details regarding the impact site and impact orientation as well as additional information about the truth simulation, please refer to the appendix.

The work presented here is similar to DART in terms of the data provided from the truth simulation, that is, a shape model for the target asteroid, the location of the impact, and many details regarding the impacting spacecraft. These pieces of information will all be known prior to or developed soon after the DART impact, and the goal of this work is to constrain the target asteroid's properties as much as possible from this limited dataset. In this study, we do not include information about the ejecta cone, anticipated from the LICIACube images \citep{dotto2021}. Thus, we are modeling what can be constrained from the minimum DART mission requirements. The exact numbers used for the inverse test, however, are not the expected values for the actual DART mission. For instance, the spacecraft used here is moving faster than expected of the DART spacecraft \citep[7.2~km/s here versus $\sim$6.1~km/s for the DART impact; ][]{stickle2022}. The extra hard hit from the spacecraft combined with the small volume of the shape model leads to a period change given for the inverse test that is almost 50\% of the orbital period of Dimorphos, a value large enough to catastrophically affect the stability of the binary system. In the actual DART impact, the velocity change for Dimorphos is expected to be a few millimeters per second or less, rather than centimeters per second. However, while the numbers and shape model used in this study differ from those expected for the DART mission, all provided data are representative of the types of information that will be available shortly after the DART impact in September 2022, and future interventions on potentially hazardous asteroids may require extreme deflections similar to this study.

\section{Spheral models}

\subsection{Spheral}

We performed our impact calculations using Spheral, an open-source Adaptive Smoothed Particle Hydrodynamics code maintained by Lawrence Livermore National Laboratory \citep[\href{https://github.com/LLNL/spheral}{https://github.com/LLNL/spheral}; ][]{owen1998,owen2010}. The simulation setup as well as two example simulations are shown in Figure \ref{fig_visit}. Spheral’s elastic--perfectly plastic formulation and approach to modeling material fracture is adapted from \citet{benz1994}. We model fracture using a tensor-based generalization of the statistical damage model of \citet{grady1980}, which accounts for the decrease in tensile strength with increasing volume. To model yield strength, we use a modified form of the of the pressure-dependent approach of \citet{collins2004} which accounts for the increase in yield strength with confining pressure. To model porous compaction, we use the strain-porosity model of \citet{wunnemann2006} with the thermal correction of \citet{collins2011}.

We simulate the impact in three spatial dimensions. The resolution of the simulations is 10~cm at the impact point and decreases outward in spherical shells, with spacing between particles increasing by a ratio of 1.01 in successive shells. This resolution equates to 3 nodes per impactor radius. Particle mass, volume, and smoothing length all vary consistent with this initial graded particle distribution and the appropriate initial density. For additional information regarding the Spheral models and material parameters used in these models, please refer to the appendix.

We made a few simplifying assumptions to define the initial conditions of our simulations. First, the impactor was modeled as a solid aluminum sphere (Figure \ref{fig_visit}b). The asteroid was modeled as a single isomorphic, homogeneous, microporous body of SiO$_2$ using an SiO$_2$ equation of state (EOS) from the Livermore Equation of State (LEOS) database \citep{fritsch2016}. The choice of SiO$_2$ as our material is not especially unusual as quartz is a well explored material in mineral physics, though we also briefly examine the use of basalt, granite, and pumice as asteroid materials, in addition to using both the ANEOS \citep{thompson1990} and Tillotson EOS \citep{tillotson1962}. Aside from the simulations testing different materials, the density of fully solid material (i.e., porosity~=~0) was held constant between simulations at 2.65~g/cm$^3$, the density of SiO$_2$ at ambient Earth conditions.

We varied seven variables describing the asteroid’s material properties: the solid yield strength of intact and damaged material ($Y_{\text{s}0}$ and $Y_{\text{d}0}$), the solid shear modulus of intact and damaged material ($G_{\text{s0}}$ and $G_{\text{d0}}$), the minimum pressure (i.e., maximum tensile pressure) allowed in intact ($P_{\text{min}}$) and damaged ($P_{\text{d,min}}$) material, and the initial porosity of the starting material ($\phi$). Properties for damaged materials were constrained to be smaller in magnitude than their intact equivalents (e.g., $Y_{\text{d}0}$~$<$~$Y_{\text{s}0}$), but otherwise parameters were allowed to vary independently. Specific ranges searched for each variable are in Table \ref{atab_search}.

\subsection{Simulation metrics}

There were three key results from each simulation: $\Delta V$ for the asteroid in the orbital direction, the momentum enhancement factor describing the contribution of the ejecta to asteroid momentum ($\beta$), and the crater morphology. Ejecta was defined as material at least 1~m away from the asteroid surface and moving faster than 5~cm/s away from the centroid of the asteroid (Figure \ref{fig_ejecta}; see appendix for more detail). $\Delta V$ was calculated from the momentum of the impacting spacecraft ($\rho_{\text{s}}$), the momentum of the ejecta ($\rho_{\text{e}}$), and the mass of the asteroid ($M_{\text{A}}$):
\begin{equation}
\Delta V = \frac{(\bm{\rho}_{\text{s}} + \bm{\rho}_{\text{e}}) \cdot \bm{\hat{o}}}{M_{\text{A}}},
\end{equation}
in which the subscripts A, s, and e refer to the asteroid, spacecraft, and ejecta, respectively. The unit vector in the orbital direction is denoted $\bm{\hat{o}}$.

The momentum enhancement factor $\beta$ is a multiplicative factor applied to the spacecraft momentum to express the contribution of impact ejecta to asteroid momentum. In its simplest form, $\beta$ is the ratio of the momentum of the asteroid after impact to the initial momentum of the impacting spacecraft. In this study, $\beta$ was calculated using the definition for momentum enhancement in the orbital direction described in \citet{rivkin2021}:
\begin{equation}
\beta = \frac{(M_{\text{A}}/M_{\text{s}})\Delta V_{\text{A,o}} - \bm{V}_{\text{s},\perp\hat{n}} + V_{\text{s,n}}\bm{\epsilon} \cdot \bm{\hat{o}}}{V_{\text{s,n}}(\bm{\hat{n}} + \bm{\epsilon}) \cdot \bm{\hat{o}}}.
\end{equation}
In this equation, $M_{\text{A}}$ is the mass of the asteroid, $M_{\text{s}}$ is the mass of the impacting spacecraft, $\bm{\hat{n}}$ is the surface normal unit vector, and $\bm{\epsilon}$ is the offset vector between $\bm{\hat{n}}$ and the velocity vector of the ejecta. $\Delta V_{\text{A,o}}$ is the change in the asteroid's velocity in the orbital direction (simply referred to as $\Delta V$ for the remainder of the paper). $V_{\text{s}}$ refers to the velocity of the impacting spacecraft and is broken down into the magnitude of velocity parallel to the surface normal for $V_{\text{s,n}}$ and the velocity vector perpendicular to the surface normal (along the surface) for $\bm{V}_{\text{s},\perp\hat{n}}$

The width ($W_{\text{c}}$) and depth ($D_{\text{c}}$) of the impact crater was measured for successful simulations (i.e., those that produced a $\Delta V$ in the range dictated by the truth simulation). Crater formation occurs over times on the order of minutes to hours rather than the fractions of a second to seconds typically run in hydrocodes. To predict the final crater as well as possible from the final time steps of our successful simulations (from 0.6~s up to 3.1~s, depending on the simulation), we applied a density constraint such that material with a density less than 95\% of the original porous density of the asteroid was assumed to eventually leave the crater. We use this metric as material in this state is typically under tension and in the process of being evacuated from the growing transient crater. For relatively small craters with diameters less than 20~m, crater sizes were calculated by fitting the cratered region of the asteroid with a plane describing the asteroid surface and a hyperboloid describing the crater \citep[Figure \ref{fig_craterFit}; ][] {klein2013}. For larger craters, crater depth and diameter were measured by hand.

\subsection{Efficiently covering search space using machine learning and extrapolation}

The large dimensional space associated with the seven input variables placed constraints on our available computational time, requiring an efficient methodology to adequately constrain the parameter space. Two main techniques were used to mitigate the computational cost (additional details for both can be found in the appendix). First, we used a machine learning decision-tree algorithm to select parameter combinations. We initialized the tree with a preliminary run of 40 simulations with randomized input parameters. For each subsequent run, between 3 and 16 simulations were chosen from 10,000 possible solutions, using the Mitchell's Best Sampling algorithm \citep{mitchell1991} to select the parameter combinations covering the largest parameter space in our seven dimensions.

Second, while the ideal case would be to run all of our simulations until stable values of $\beta$ and $\Delta V$ were reached, as well as stable crater formation, the size of our dataset makes this approach prohibitively computationally expensive. Thus, as the truth model provided a relatively narrow target range for $\Delta V$ of $\pm$0.07~cm/s, simulations that clearly overshot or undershot the velocity range were terminated prior to $\Delta V$ and $\beta$ stabilizing. The relationship between $\Delta V$ and time was extrapolated using an exponential decay model, taking the infinite time limit as the stable $\Delta V$ (see Figure \ref{fig_extrap} for all simulation fits; see appendix for more detail). For simulations with high porosity, a local maximum in $\Delta V$ was often observed at small times (t~$\sim$~0.1--0.3 s). For the purposes of extrapolation, this local maximum was fit phenomenologically by adding a Gaussian peak to the exponential decay model. $\beta$ was extrapolated in the same way, with or without an additional Gaussian peak depending on the porosity in the simulation. Simulations with values of $\Delta V$ close to the target $\Delta V$ were run until relatively stabilized (i.e., with a small $\delta V/\delta t$).

\section{Results} \label{sec:results}

\subsection{Trends across all simulations}

\begin{figure*}
\fig{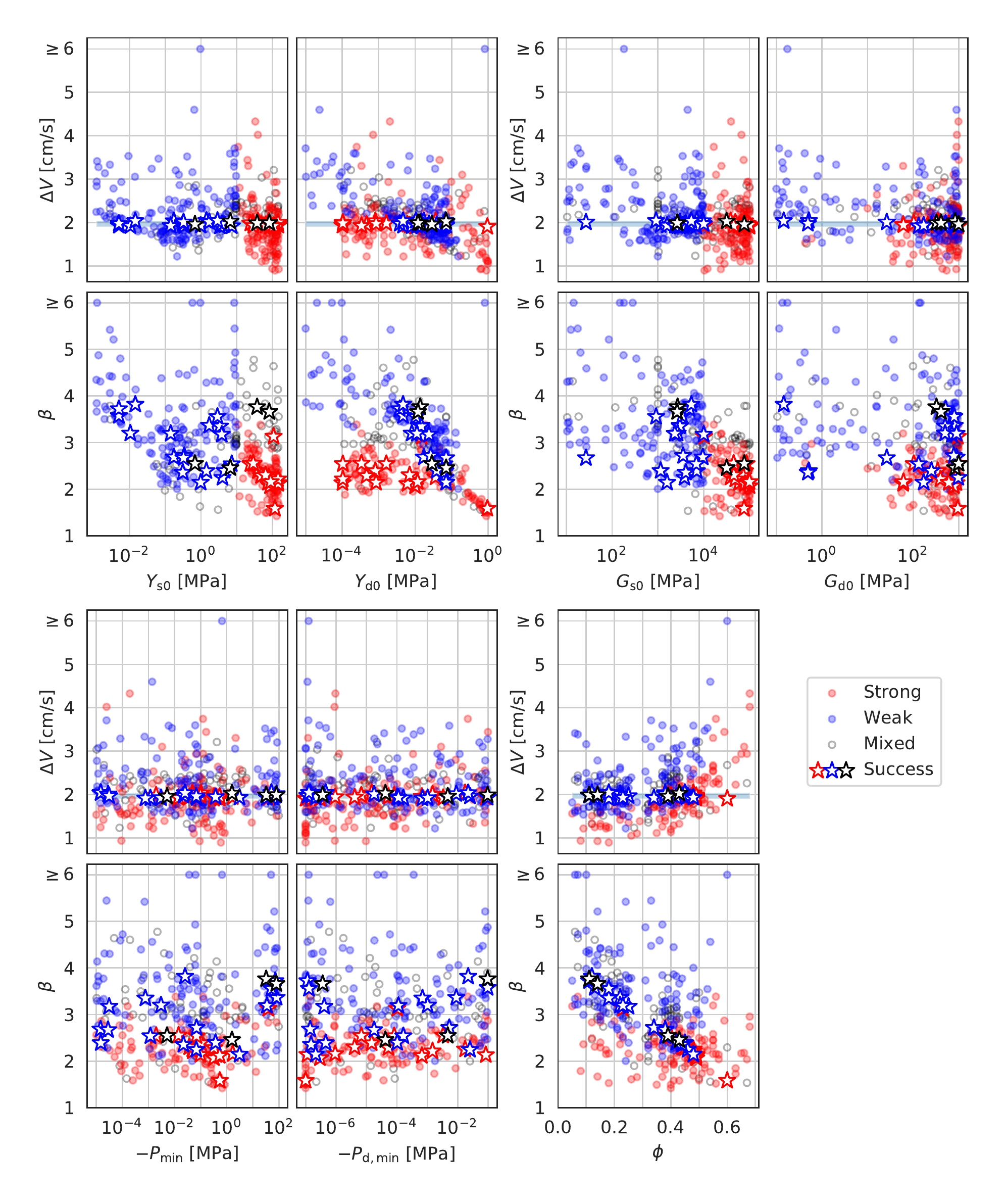}{0.9\textwidth}{}
\caption{Results of all simulations. Final $\Delta V$ and $\beta$ are extrapolated to $t = \infty$. Simulations with $Y_{\text{s0}} > 10$ MPa and $G_{\text{s0}} > 10$ GPa are colored in red and are more consistent with strong, intact rock. Simulations with $Y_{\text{s0}} < 10$ MPa and $G_{\text{s0}} < 10$ GPa are colored in blue and are more consistent with weak, fractured rock or sediment. Simulations in open grey circles have either a large $Y_{\text{s0}}$ and a low $G_{\text{s0}}$, or a low $Y_{\text{s0}}$ and a large $G_{\text{s0}}$. Simulations with final $\Delta V$ values in the target range are denoted with stars. Simulations with $P_{\text{d,min}} = 0$ are plotted at $10^{-7}$ MPa.}
\label{fig_weakstrong}
\end{figure*}

We ran 338 simulations covering a wide range of input parameters. The results of these simulations are summarized in Figure~\ref{fig_weakstrong}, the ranges of inputs and outputs covered are listed in Table~\ref{tab_res}, and all inputs and deflection results can be found in Table \ref{atab_alldata}. For clarity, we plot $\Delta V$ as its magnitude, but all values of $\Delta V$ are negative in our simulations (i.e., the asteroid orbital period decreases after impact).

\begin{deluxetable*}{lllllllll}[t]
\tablecaption{Ranges of input and output values for truth simulations and inverse simulations}
\tablehead{
& \multicolumn{2}{l}{Truth simulation} & \multicolumn{2}{l}{All inverse simulations} & \multicolumn{2}{l}{Simulation subset} & \multicolumn{2}{l}{Simulation subset}\\
& & & \multicolumn{2}{l}{} & \multicolumn{2}{l}{$\Delta V = -1.972 \pm 0.07$ cm/s} & \multicolumn{2}{l}{$\Delta V = -1.972 \pm 0.07$ cm/s}\\
& CTH & FLAG & \multicolumn{2}{l}{} & \multicolumn{2}{l}{} & \multicolumn{2}{l}{$M_{\text{A}} = 599,758 \pm 59,976$ kg}\\
& & & \multicolumn{2}{l}{} & \multicolumn{2}{l}{} & \multicolumn{2}{l}{~~~($\phi = 0.18 \pm 0.08$)}
}
\startdata
$n$ & 1 & 1 & \multicolumn{2}{l}{338} & \multicolumn{2}{l}{37} & \multicolumn{2}{l}{12}\\
\hline
& & & \textbf{\underline{Minimum}} & \textbf{\underline{Maximum}} & \textbf{\underline{Minimum}} & \textbf{\underline{Maximum}} & \textbf{\underline{Minimum}} & \textbf{\underline{Maximum}} \\
$Y_{\text{s}0}$ [MPa] 	& 1.00e+1 	& 1.00e+1 	& 1.00e$-$3 	& 1.50e$+$2 	& 5.01e$-$3 	& 1.46e$+$2 	& 5.01e$-$3 	& 1.07e$+$2 \\
$Y_{\text{d}0}$ [MPa] 	& --- 		& --- 		& 1.00e$-$5 	& 1.00e$+$0 	& 1.02e$-$3 	& 9.52e$-$1 	& 3.80e$-$3 	& 6.92e$-$2 \\
$G_{\text{s0}}$ [MPa] 	& ---\tablenotemark{a}		& 2.90e+3 	& 1.00e$+$1 	& 1.00e$+$5 	& 1.07e$+$1 	& 1.00e$+$5 	& 1.07e$+$1 	& 1.02e$+$4 \\
$G_{\text{d0}}$ [MPa] 	& --- 		& --- 		& 1.00e$-$1 	& 1.00e$+$3		& 1.45e$-$1 	& 9.91e$+$2 	& 1.45e$-$1 	& 9.55e$+$2 \\
$\phi$ 					& 0.24 		& 0.24 		& 0.05 			& 0.70 			& 0.11 			& 0.6 			& 0.11 			& 0.25 \\
$-P_{\text{min}}$ [MPa] & --- 		& --- 		& 1.00e$-$5 	& 1.00e$+$2 	& 1.45e$-$5 	& 8.13e$+$1 	& 3.09e$-$5 	& 8.13e$+$1 \\
$-P_{\text{d,min}}$ [MPa] & --- 	& --- 		& 0 			& 1.00e$-$4 	& 0 			& 9.55e$-$2 	& 1.15e$-$7 	& 9.55e$-$2 \\
\\
$\Delta V$ [cm/s] & $-$2.02 & $-$1.92 & $-$0.89 & $-$4.60\tablenotemark{b} & $-$1.90 & $-$2.04 & $-$1.91 & $-$2.04 \\
$\beta$ & 3.3 & 3.5 & 1.4 & 6.9\tablenotemark{b} & 1.53 & 3.59 & 2.88 & 3.59 \\
$W_{\text{c}}$ [m] & 9.2\tablenotemark{c} & --- & --- & --- & 11.3 & 47.0 & 19.2 & 47.0 \\
$D_{\text{c}}$ [m] & 6.9\tablenotemark{c} & --- & --- & --- & 3.6 & 15.9 & 4.9 & 14.4 \\
\enddata
\tablecomments{The truth simulation parameters were unknown to the inverse team when inverse simulations were being run, but they are included here for ease of comparison. The total inverse simulation population is broken down into subsets meeting specific constraints. For each subset, the criteria for selection are listed in the header and the corresponding range of material properties are shown. Crater width and crater depth were only calculated for simulations with the correct $\Delta V$, so ranges are not reported for the whole simulation set.}
\tablenotetext{a}{While an intact shear modulus was not set in the CTH simulation, the Poisson's ratio was set to 0.25.}
\tablenotetext{b}{Does not include values for simulation with fluid-like behavior.}
\tablenotetext{c}{Transient crater measured at 0.1~s.}
\label{tab_res}
\end{deluxetable*}

\vspace{-2.5em}

One of the largest individual effects on $\Delta V$ comes from porosity, with larger values of $\Delta V$ associated with more porous targets (Figure~\ref{fig_weakstrong}). This observation is predominantly a result of the effect of porosity on the total initial mass of the asteroid, as asteroid volume and solid density (i.e., density at 0\% porosity) were constant between simulations. The momentum of the impacting spacecraft is constant between simulations. Thus, neglecting any effect of impact ejecta, conservation of momentum results in the velocity change of the asteroid is larger when the asteroid has less mass. This effect can be observed in the positive relationship between porosity and $\Delta V$, particularly for high values of porosity.

Strength and shear modulus for both intact and damaged material have negative correlations with $\Delta V$ (Figure~\ref{fig_weakstrong}), meaning that simulations with yield strengths and shear moduli that more closely approximate fully intact rock are more likely to result in small velocity changes. At the other extreme, simulations that resemble weak dry sediment or sand are more likely to have large velocity changes. Previous work examining the effects of yield strength or cohesion on $\Delta V$ and $\beta$ also observed these correlations \citep[e.g.,][]{brucksyal2016,raducan2019}.

$P_{\text{min}}$ and $P_{\text{d,min}}$ do not have strong effects on $\Delta V$ across the simulations. There is essentially no correlation between the magnitude of $P_{\text{min}}$ and $\Delta V$ and only a shallow positive correlation between the magnitude of $P_{\text{d,min}}$ and $\Delta V$ across six orders of magnitude (Figure~\ref{fig_weakstrong}). To quantify these relationships (or lack thereof), we ran a Monte Carlo regression model, in which we fit a regression to a subset of our dataset 10,000 times and took the average and standard deviation of those 10,000 results. For the trends of both $P_{\text{min}}$ and $P_{\text{d,min}}$ versus $\Delta V$, the standard deviation of the regression results for the slope is larger than the average slope, suggesting that any trend observed is statistically insignificant. The general lack of a trend between $P_{\text{min}}$ and $P_{\text{d,min}}$ and the resulting deflection suggests that tensile stresses in our simulations are not controlling the overall deflection response of the asteroid.

\begin{figure}
\fig{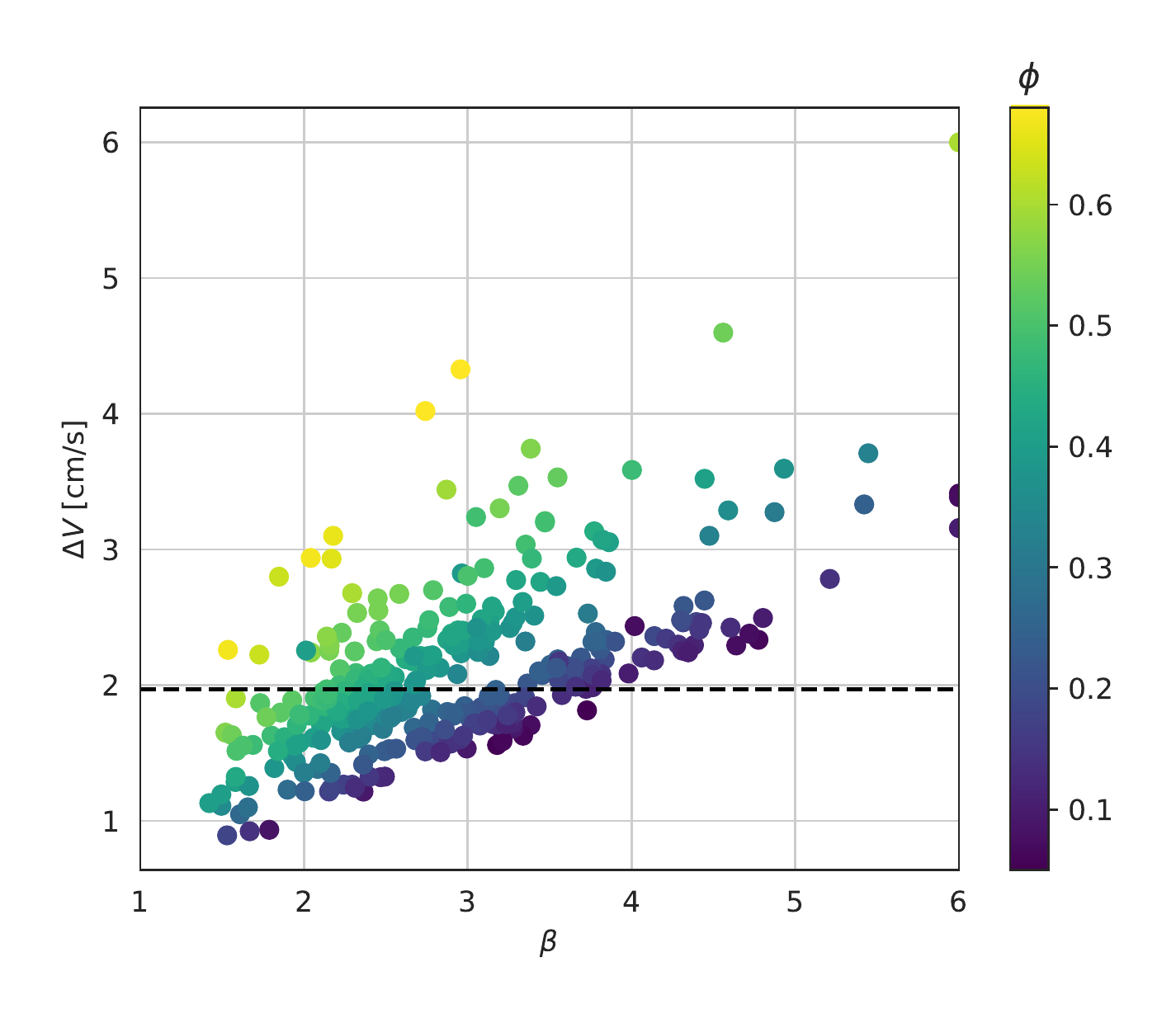}{0.48\textwidth}{}
\caption{Final $\Delta V$ versus $\beta$ for all simulations. Each simulation is colored by initial porosity, $\phi$. The target $\Delta V$ of $-$1.972~cm/s is indicated with a dashed line.}
\label{fig_delVbeta}
\end{figure}

As the mass of the asteroid is not constant between simulations, there is not a linear relationship between $\Delta V$ and $\beta$ (Figure~\ref{fig_delVbeta}), but the two are positively correlated through the general equation for $\beta$ (Eq. 2). In fact, for most variables, the trends observed with $\Delta V$ and $\beta$ are the same, i.e., both positive, both negative, or both essentially negligible (Figure~\ref{fig_weakstrong}). The one exception is porosity. The correlation between $\phi$ and $\Delta V$ is positive as high-porosity targets have lower mass. However, there is a negative relationship between porosity and $\beta$. In more porous targets, a larger portion of the kinetic energy of the impactor goes toward compaction of the target material rather than the ejection of material. These results are consistent with previous work examining the effects of porosity on asteroid deflection \citep[e.g.,][]{brucksyal2016,raducan2019,stickle2017}. 

\subsection{Defining realistic targets}

Our dataset covers a large range of properties, and it is worth emphasizing that we did little to control the realism of material parameter combinations. While this methodology allows us to easily and effectively explore the parameter space, some simulations are not predictive for real asteroids that might be encountered. It is thus instructive to examine trends in subsets of data that are more representative of real material.

\subsubsection{Strong and weak targets}

Figure~\ref{fig_weakstrong} uses the intact asteroid strength and shear modulus to define realistic combinations representing strong and weak target material. Simulations shown in red in Figure~\ref{fig_weakstrong} are a subset with large intact strength and shear moduli more consistent with rock ($Y_{\text{s}0}$~$>$~10~MPa, $G_{\text{s0}}$~$>$~10~GPa). Simulations in blue represent weaker material more consistent with fractured rock or sediment ($Y_{\text{s}0}$~$<$~10~MPa, $G_{\text{s0}}$~$<$~10~GPa). Finally, simulations in white are simulations in which one intact parameter is large while the other intact parameter is small (e.g., large $Y_{\text{s}0}$ and small $G_{\text{s0}}$), combinations which are unlikely to occur. We do not include porosity in this comparison.

In both the strong and weak subsets, the trends between porosity and $\Delta V$ are much clearer for intermediate and very porous materials ($\phi$~$>$~20\%) than the trend for the whole dataset. For the same porosity, weak target material generally results in a larger $\Delta V$ compared to stronger target material, aligning with the overall trends observed for the full dataset (Figure~\ref{fig_weakstrong}). Similarly, weak target material generally results in a larger $\beta$ for the same porosity.

For the initially strong, more rock-like simulations (in red), there is not a strong trend between $\Delta V$ and the variables for damaged material ($Y_{\text{d}0}$ and $G_{\text{d0}}$) (Figure~\ref{fig_weakstrong}). The lack of a trend is particularly evident for $Y_{\text{d}0}$: the simulations in red at low $Y_{\text{d}0}$ cover ranges of values for $\Delta V$ and $\beta$ similar to simulations at high $Y_{\text{d}0}$, particularly for $Y_{\text{d}0} < 10^{-1}$~MPa. In contrast, when $Y_{\text{s}0}$ and $G_{\text{s0}}$ are both small, $\Delta V$ and the damaged variables ($Y_{\text{d}0}$ and $G_{\text{d}0}$) have strong negative correlations. These relationships suggest that the ejecta response for strong asteroid targets is controlled by the initial material parameters ($Y_{\text{s}0}$ and $G_{\text{s0}}$), and the initial yielding of asteroid material is the limiting step in ejecta formation. For initially weak asteroid targets, on the other hand, the ejecta response is controlled by the parameters for damaged material ($Y_{\text{d}0}$ and $G_{\text{d0}}$), and the deformation of material that is already damaged controls the ejecta formation.

\subsubsection{Realistic elastic parameters}

We searched a large range of shear moduli, but we note that we did not directly link the bulk modulus to the prescribed shear modulus. Instead, the bulk modulus for each simulation was calculated from the SiO$_2$ EOS. As a result, the Poisson’s ratio ($\nu$) of our simulations varied substantially. Simulations with low shear moduli had very large Poisson’s ratios, close to the maximum value of 0.5. By comparison, porous rhyolitic lavas (up to $\phi$~=~0.5) also demonstrate Poisson’s ratios of up to $\sim$0.45 \citep[e.g.,][]{ji2019,mordensky2018}. Vacuum-saturated beach sands demonstrate a large range of Poisson’s ratios but have a maximum of 0.42 for typical grain sizes of hundreds of micrometers to around a millimeter \citep{kimura2006}. Poisson’s ratios larger than $\sim$0.45 in geologic material, however, are uncommon without introducing pore fluids \citep[e.g.,][]{kimura2006}, which are unrealistic for asteroid environments.

At the other extreme, simulations with extremely large shear moduli require auxetic behavior with negative Poisson’s ratios. Auxetic materials generally have very low densities resulting from their complex open structures \citep[e.g.,][]{alderson2007}, and they are exceedingly rare in geologic materials. Alpha-cristobalite, a high-temperature polymorph of SiO$_2$, is the only known natural mineral with a negative Poisson’s ratio \citep{ji2018}, though some artificial pumice analogues have been reported to have negative Poisson’s ratios as well \citep{wollner2018}.

Examining only the simulations with solid Poisson’s ratios similar to typical geologic material (0--0.45), the trends we observe across simulations do not change (Figure~\ref{fig_poisson}). The ranges of material parameters for successful simulations are similarly unaffected, aside from the solid shear modulus as it is directly constrained by the Poisson’s ratio. Given the generally unchanged results, we will consider all simulations, including those with extreme Poisson’s ratios, in the remainder of the manuscript.

\section{Predicting asteroid properties} \label{sec:discussion}

\subsection{Matching $\Delta V$}

\begin{figure*}
\fig{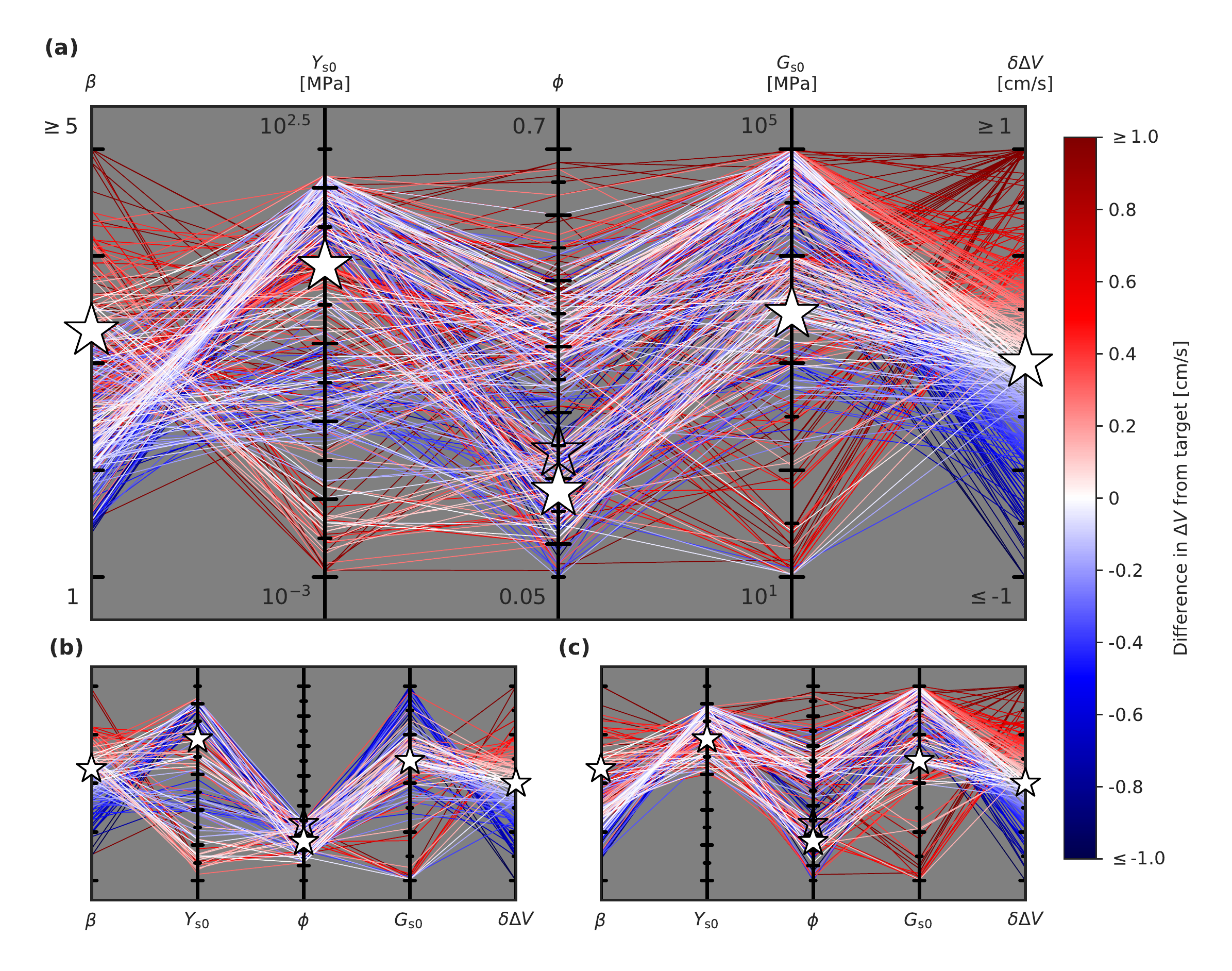}{\textwidth}{}
\caption{Parallel coordinate plots of input parameters with equivalent parameters in the truth simulations, as well as final $\Delta V$ and $\beta$. Simulation lines are colored by the difference between the simulation $\Delta V$ and the target $\Delta V$, with blue being slower than the truth simulation and red being faster than the truth simulation. White stars indicate the values of the truth simulation. For $\phi$, the open star indicates the original truth simulation value of 0.24 while the solid white star indicates the porosity required for an equivalently massive asteroid in our simulations. Plots are shown for (a) all simulations, (b) simulations with initial asteroid mass within 10\% of the truth simulation, and (c) simulations with $Y_{\text{s0}}$ within one order of magnitude of the truth simulation. Ticks and axis limits in (b) and (c) are the same as in (a).}
\label{fig_spag-main}
\end{figure*}

Thus far, we have discussed the trends observed across all our simulations. However, the goal of the simulation efforts presented here is to find material parameter combinations that produce a particular $\Delta V$, in this case $-$1.972~cm/s. These solutions are shown as stars in Figure~\ref{fig_weakstrong} and are white lines in Figure~\ref{fig_spag-main}. The minimum and maximum values for each variable for successful simulations (i.e., those simulations with $\Delta V$~=~$-$1.972~$\pm$~0.07~cm/s) are reported in Table~\ref{tab_res}. Example visualizations of simulations that produced the correct $\Delta V$ with very different material parameters are shown in Figure \ref{fig_visit}c--d.

Looking at material parameters in isolation, the successful simulations cover a large range of properties. In line with the small effect of these variables on $\Delta V$, both $P_{\text{min}}$ and $P_{\text{d,min}}$ for successful simulations cover nearly the entire range of search space. Porosities for successful simulations are in the range of 11\% to 60\%. This range covers everything from roughly intact rock up to the porosities observed in rubble-pile asteroids like Itokawa or 101955 Bennu \citep{walsh2018}. The large range in asteroid mass associated with these porosities leads to a range of values for $\beta$ from 1.5 to 3.6.

Maximum values for $Y_{\text{s}0}$ and $G_{\text{s0}}$ for these simulations are representative of strong rock ($Y_{\text{s}0}$~=~146~MPa, $G_{\text{s0}}$~=~100~GPa). Minimum values ($Y_{\text{s}0}$~=~174~kPa, $G_{\text{s0}}$~=~27~MPa) are similar to the cohesion and shear modulus for clay \citep{onur2014}. The strength and shear modulus for damaged material in successful simulations are also similar to the properties of sediment. $Y_{\text{d}0}$ varies from 100~Pa to 71~kPa, covering a range similar to that expected for lunar regolith \citep{holsapple2012} or the surface of the rubble-pile asteroid 162173 Ryugu \citep{arakawa2020}. With the exception of two simulations with $G_{\text{d0}}$ of $\sim$500~kPa and one simulation with $G_{\text{d0}}$ of $\sim$150~kPa, the shear moduli for damaged material range from $\sim$30 MPa to $\sim$1 GPa, demonstrating elasticity similar to sand on the low end of that range \citep{onur2014}. Shear moduli of $\le$500~kPa, however, are quite small for geologic material, and nearly all the simulations with similar shear moduli for damaged material result in velocity magnitudes larger than the target $\Delta V$. One simulation with $G_{\text{d0}}$ $\sim$200~kPa, in addition to low strength and $G_{\text{s0}}$, even exhibited fluid-like behavior (Figure~\ref{fig_fluid}). While examining low values for $G_{\text{d0}}$ was helpful for establishing the algorithmic search space, this behavior is extremely unlikely for actual asteroid material.

\subsection{Reproducing the truth simulation material parameters}

The results from the truth simulation were defined using the average of results from two different codes: CTH \citep{mcglaun1990} and Free LAGrange \citep[FLAG;][]{burton1992,burton1994a,burton1994b}. These simulations were run to a final time of 0.1~s in CTH and 0.2~s in FLAG. Both simulations used a basalt EOS, SESAME \citep{lyon1992} for CTH or Mie-Gr\"{u}neisen \citep{meyers1994} for FLAG, to describe the asteroid material, with a yield strength of 10 MPa, a density of 2.855~g/cm$^3$ for fully solid material, and a porosity of 24\% (and thus a bulk density of 2.170~g/cm$^3$). The FLAG simulation explicitly set a shear modulus for intact material of 2.9~GPa while the CTH simulation set a Poisson's ratio of 0.25. The average $\beta$ resulting from the CTH and FLAG truth simulations was 3.4. These parameters are listed in Table \ref{tab_res}, and additional detail about the truth simulation codes can be found in the appendix. Figure~\ref{fig_spag-main} compares the inverse simulations to the truth simulation in a parallel coordinate plot, with the truth simulation parameters plotted as stars. (For a parallel coordinate plot of all of the input parameters used in the Spheral simulations, see Figure \ref{fig_spaghettiall} in the appendix.)

With the wide range of parameters covered by our successful simulations, the truth simulation parameters fall inside the ranges described in the previous section (Figure~\ref{fig_spag-main}a). Indeed, several of our parameter combinations are quite close to the truth simulation. However, solutions with similar parameters are uncommon overall. For instance, values of $Y_{\text{s}0}$ within one order of magnitude of the truth simulation value of 10~MPa are associated with both low and high porosities, with higher porosities generally correlated with larger shear moduli (Figure~\ref{fig_spag-main}c). Looking at the inverse correlation, successful simulations with low porosities, and thus asteroid masses similar to the truth simulation, are associated with both high strengths in the range of MPa and low strengths of a few kPa (Figure~\ref{fig_spag-main}b).

Beyond the multi-solution degeneracy of the problem of asteroid deflection, there are several potential explanations for why the parameter combination from the truth simulation is uncommon in our inverse simulations. First, there are inherent differences in the way different hydrocodes process impact problems. Each code in this study utilizes a different discretization method: CTH is an Eulerian finite-difference code, FLAG is an Arbitrary Lagrange-Eulerian finite-volume code, and Spheral is a smoothed particle hydrodynamics code. There are also other implementation aspects that differ between the three codes. For example, once material is fully damaged in both the CTH and FLAG simulations, it can no longer support any stress. Conversely, the Spheral simulations explicitly set both a strength ($Y_{\text{d}0}$) and a shear modulus ($G_{\text{d0}}$) for fully damaged material \citep[see][for additional details regarding the damage model in Spheral]{owen2022}. Thus, Spheral simulations with the same parameters describing intact material as the truth simulation (i.e., $Y_{\text{s}0}$, $G_{\text{s0}}$, and $\phi$) could have smaller values for $\Delta V$ and $\beta$ due to the contribution of non-zero values of $Y_{\text{d}0}$ and $G_{\text{d0}}$.
Even when great effort is made to make sure hydrocodes are working on as identical a problem as possible (e.g., by constraining material parameters, equations of state, and strength models), there are still differences in final simulation results with variations in $\beta$ of 15--20\% between the different codes \citep{stickle2020}. %

For the inverse test, we used the differences between codes to model the imperfections of using our hydrocodes to simulate reality. As the truth simulation was only a model itself, however, we can further examine the CTH, FLAG, and Spheral simulations to pull apart some of the more minute differences. For instance, in benchmarking tests using a basalt EOS to describe the target, CTH predicts larger values of $\beta$ (and thus larger values of $\Delta V$) than Spheral for targets with the same properties \citep{stickle2020}. Applying this difference to the inverse problem described in this study suggests that replicating the truth simulation parameters in Spheral would result in a smaller $\beta$ and $\Delta V$ than the CTH results. Thus, in order to reproduce the correct $\Delta V$ with an asteroid mass close to the truth simulation, more of our successful simulations have lower values of $Y_{\text{s}0}$ (Figure~\ref{fig_spag-main}b). Conversely, many successful simulations with $Y_{\text{s}0}$ close to 10~MPa have higher porosities than the truth simulation (Figure~\ref{fig_spag-main}c).

\begin{figure}
\fig{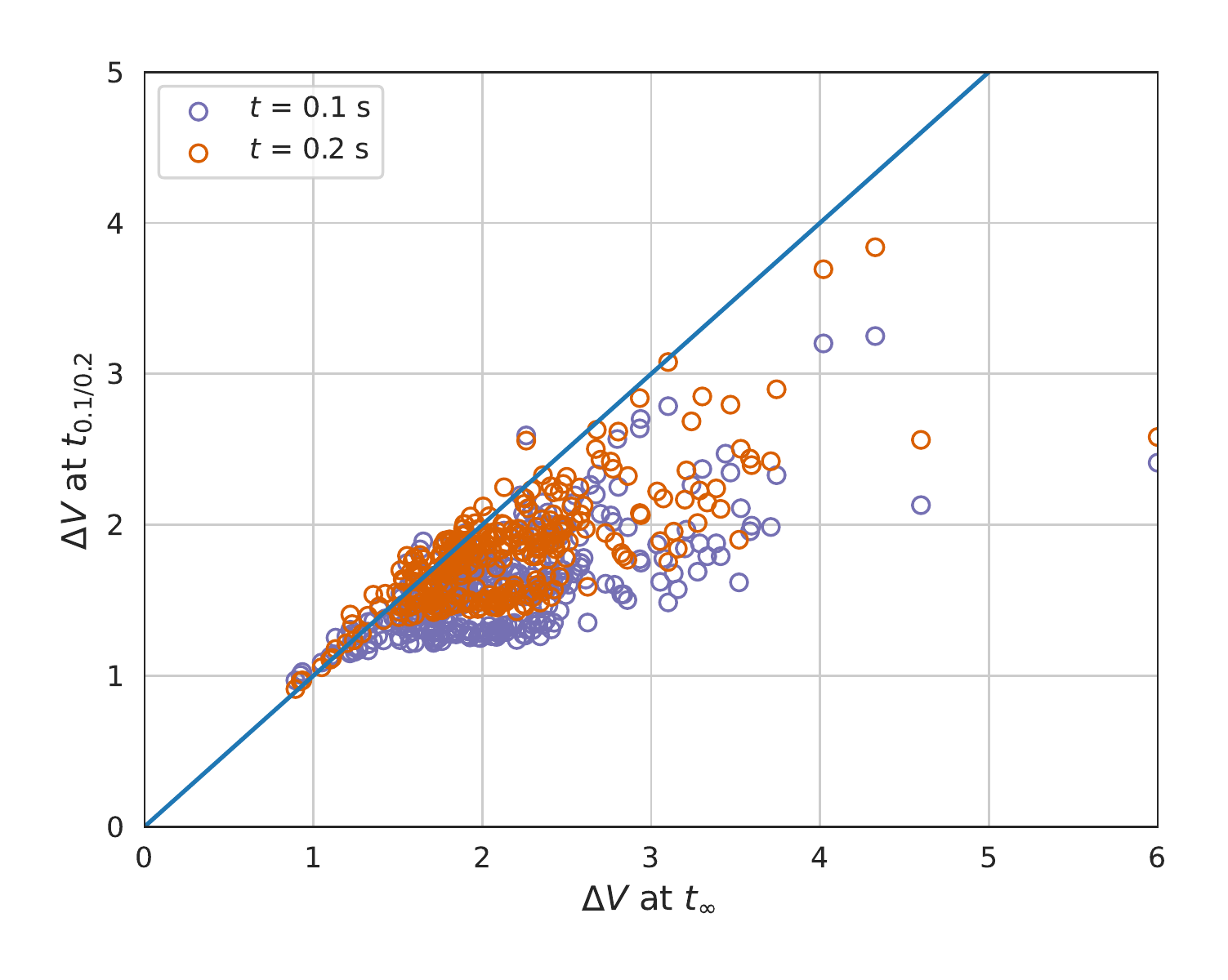}{0.48\textwidth}{}
\caption{$\Delta V$ at $t = 0.1$~s (purple) or $t = 0.2$~s (orange) versus $\Delta V$ at $t \sim \infty$. The blue line is 1:1.}
\label{fig_simtime}
\end{figure}

We also consider the simulation time at which $\Delta V$ is measured or estimated. The truth simulation reported $\Delta V$ at $t$~=~0.1~s and 0.2~s after impact, while we extrapolate $\Delta V$ to $t \sim \infty$. If we compare the $\Delta V$ at 0.1~s ($\Delta V_{0.1}$) with the extrapolated infinite-time $\Delta V$ ($\Delta V_{\infty}$) for our simulations, the extrapolated $\Delta V$ is nearly always larger than $\Delta V_{0.1}$ (Figure~\ref{fig_simtime}). This observation also holds for $\Delta V$ measured at 0.2~s ($\Delta V_{0.2}$). The exceptions are generally simulations for strong targets, where a local maximum is observed in $\Delta V$ at early times. Since $\Delta V_{\infty}$ was larger than $\Delta V_{0.1}$ for most simulations, simulations with $\Delta V_{0.1} = -1.972\pm0.07$ cm/s were generally shifted to weaker targets (i.e., targets with larger $\Delta V_{\infty}$) compared to simulations in which $\Delta V_{\infty}$ was in the target velocity range. 

An additional discrepancy between the truth simulation and our Spheral simulations was the choice of EOS. Our choice of LEOS SiO$_2$ for the asteroid material does not match the SESAME or Mie-Gr\"{u}neisen basalt EOS used in the truth simulation. To briefly explore the effect of EOS and material choice, we chose four of our successful simulations covering a range of material property combinations. We reran these simulations using different choices for asteroid material and EOS while keeping the seven input material properties (i.e., $Y_{\text{s}0}$, $Y_{\text{d}0}$, etc.) the same. The EOS/material combinations we tested were Tillotson basalt, Tillotson granite, Tillotson pumice, and ANEOS SiO$_2$ \citep{thompson1990,tillotson1962}. 

\begin{figure*}
\fig{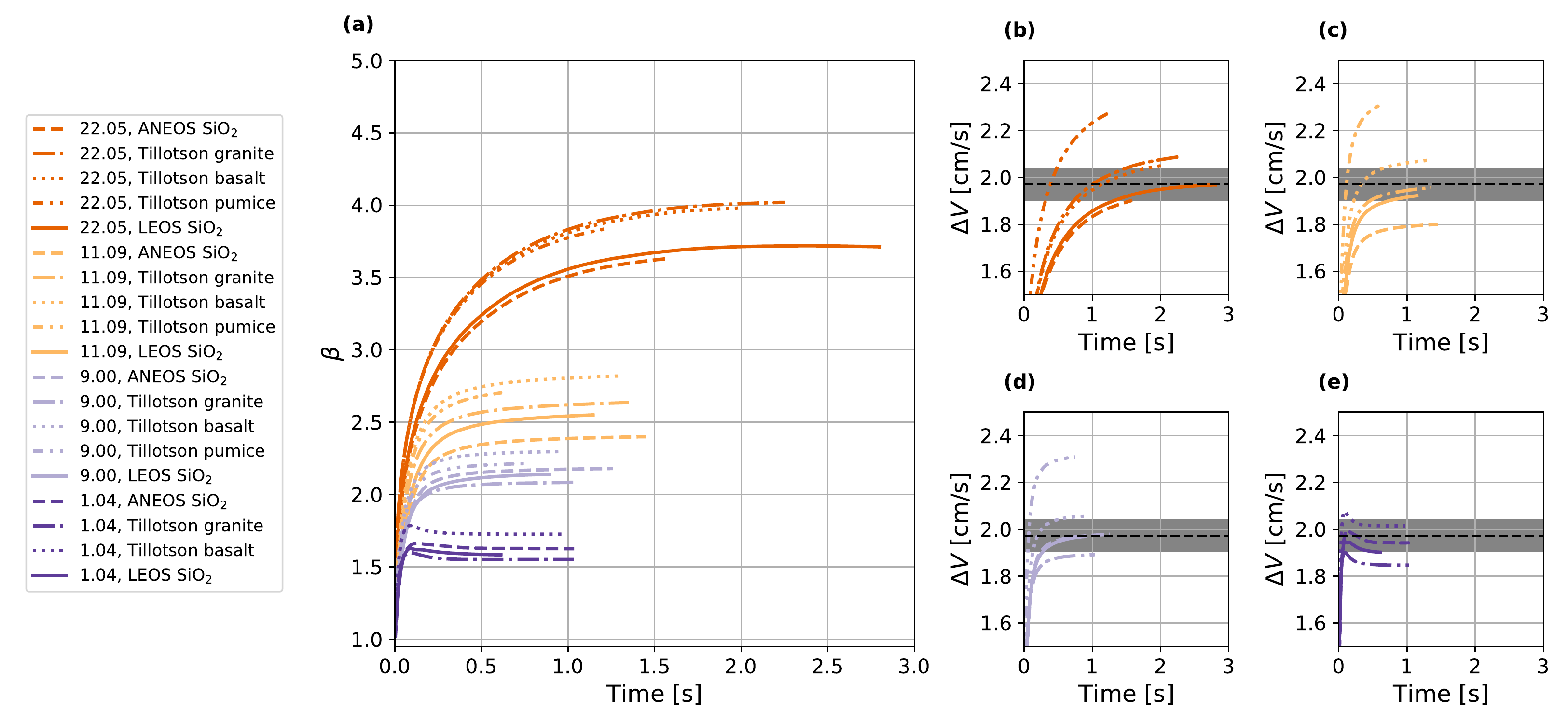}{0.9\textwidth}{}
\caption{Results of simulations examining the effect of EOS/material choice showing (a) $\beta$ versus time, and (b--e) $\Delta V$ versus time. Lines are colored to group simulations with the same material input parameters, and line type indicates the EOS and material choice. The dashed black lines in plots (b) through (e) indicates the $\Delta V$ of the truth simulation, with the dark grey covering $\pm$0.07~cm/s. Numbers in the legend (e.g., 22.05) are simulation IDs (see Table \ref{atab_alldata}).}
\label{fig_EOS}
\end{figure*}

Variability in $\beta$ resulting from the choice of EOS and material for the asteroid is 10--16\% (Figure~\ref{fig_EOS}, Table~\ref{atab_EOS}), consistent with previous observations \citep{brucksyal2016,stickle2017}. Variability in $\Delta V$ is much larger (up to 26\%), predominantly because of the varying nonporous densities of the different EOS and material options. However, when only considering the SiO$_2$, basalt, and granite EOS options, which all have similar nonporous densities, variability in $\Delta V$ is 8–14\%, comparable to the variability in $\beta$. The LEOS and ANEOS SiO$_2$ results are similar to each other both in terms of $\beta$ as well as $\Delta V$, and the Tillotson EOS have similar values of $\beta$ (Figure~\ref{fig_EOS}). Interestingly, the different asteroid EOS and material combinations are not consistent in their distribution. For instance, using Tillotson granite to describe the asteroid resulted in the smallest $\Delta V$ and $\beta$ for the two strongest cases tested here (1-4 and 9-0), but ANEOS SiO$_2$ had the smallest $\Delta V$ and $\beta$ for the two weakest cases (22-5 and 11-9).

\subsection{Constraints from Hera after impact}

Additional constraints for the DART impact will be provided by Hera when it reaches the Didymos system in 2026. Hera will measure the mass of Dimorphos, with an expected accuracy of at least 10\% \citep{michel2018}. Figure~\ref{fig_spag-main}b demonstrates that we have three main populations of values for $Y_{\text{s}0}$ when constraining the mass of the target asteroid: tens of MPa, a few MPa, or a few kPa. Remarkably, however, the shear modulus for all successful mass-constrained simulations is within approximately half an order of magnitude of the truth simulation. The ranges for different input variables of successful solutions when constrained by asteroid mass are in Table~\ref{tab_res}.

\begin{figure*}
\fig{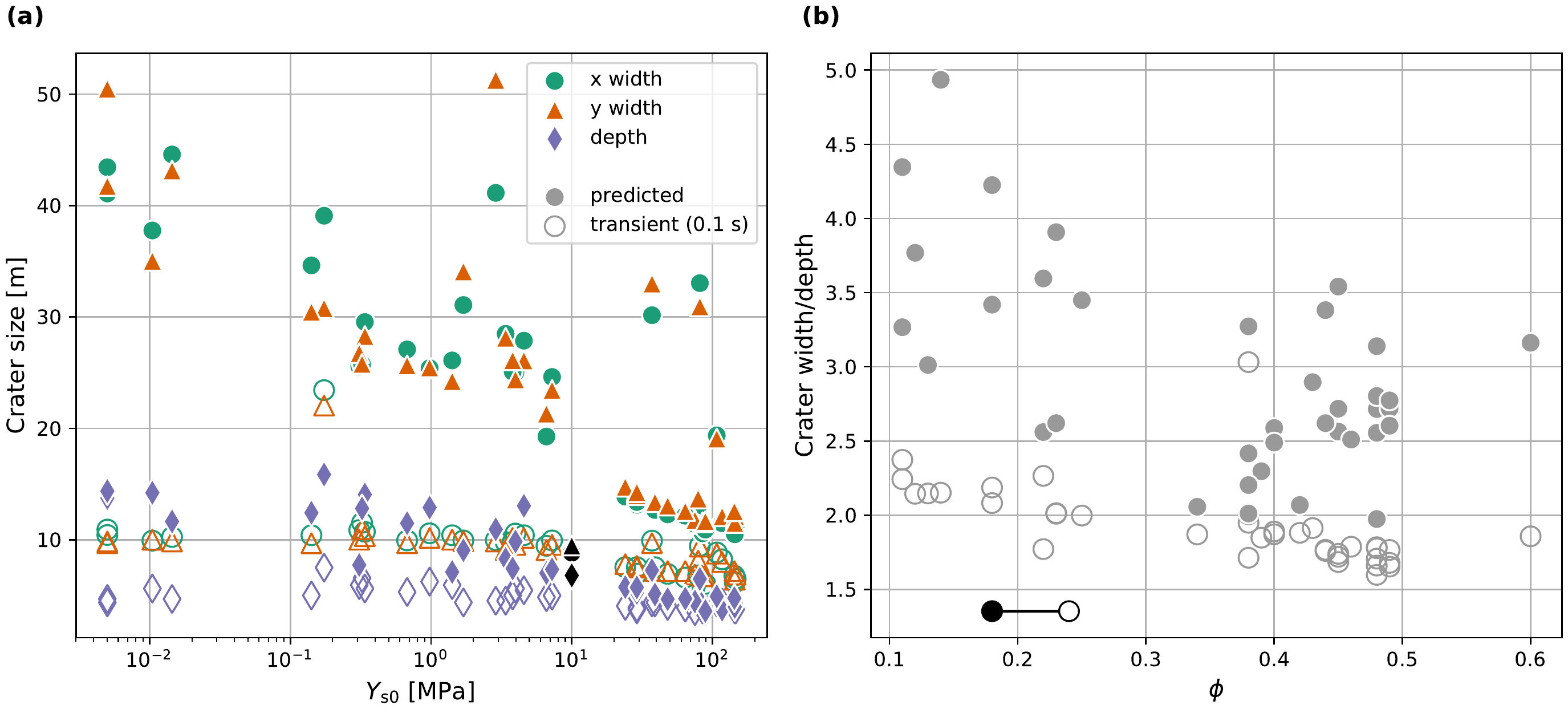}{0.9\textwidth}{}
\caption{(a) Crater size versus $Y_{\text{s0}}$ for successful simulations. Solid symbols represent the crater estimated from the final simulated time. Empty symbols represent the crater estimated at 0.1~s. The truth simulation is shown in black. (b) The ratio of crater width to crater depth versus $\phi$. Solid and empty symbols have the same interpretation as in (a). The truth simulation is shown in black, with an empty symbol for the 24\% porosity set for the basalt EOS of the truth simulation and a solid symbol for the 18\% porosity required for an SiO$_2$ EOS to have the same bulk density.}
\label{fig_crater}
\end{figure*}

In addition to mass, Hera will also measure the size of the impact crater left by DART. The morphology of the residual crater after an impact is expected to vary substantially with material properties \citep[e.g.,][]{holsapple2012}, and we observe a wide variety of crater sizes among successful simulations (Figure~\ref{fig_crater}, Table~\ref{atab_crater}). In general, both crater width and crater depth are inversely correlated with $Y_{\text{s}0}$ (Figure~\ref{fig_crater}a). The crater width in the x direction and the crater width in the y direction are typically close, though not the same. Small craters generally have slightly wider widths in the y direction, likely due to the mild obliquity of the impact interacting with the topography of the impact location. For larger craters, there are examples with the width in the x direction being larger as well as examples with width in the y direction being larger. This variation is at least in part a result of the complicated crater shapes predicted for these simulations. In addition, for a given $Y_{\text{s}0}$ (or narrow range of values for $Y_{\text{s}0}$), crater width decreases with increasing porosity. The ratio of crater width to crater depth is also correlated with porosity, with impacts into more porous asteroids generating craters with lower aspect ratios (Figure~\ref{fig_crater}b).

The crater of the CTH truth simulation was measured at 0.1~s (the final time step of that simulation) and had a width of 9.2~m and a depth of 6.9~m. While the crater depth of the truth simulation falls directly on the general trend observed in our simulations, the crater width of the truth simulation is significantly smaller. This difference is observable both in the general trend of crater width versus $Y_{\text{s}0}$ as well as crater size ratio versus $\phi$ (Figure \ref{fig_crater}). The underestimation of crater size relative to the Spheral simulations likely results from the transient nature of the crater in the CTH simulation. When measuring crater size in the Spheral simulations, we specifically attempted to predict a late-time crater shape by removing material below a critical density after $\Delta V$ stabilized. For the majority of crater shapes in our simulations, the change in width between the transient crater and the predicted final crater is larger than the change in depth. Thus, we expect that the predicted final crater in the truth simulation would similarly demonstrate a large increase in crater width relative to the transient with only a small increase in crater depth, bringing the truth simulation crater size into relative alignment with the trends we observe.

Given the transient nature of the truth simulation crater, it is difficult to use crater size as a precise constraint on our simulations (compared to the easily quantified asteroid mass, for instance). In the interest of exploring how crater size might narrow the range of possible material properties, however, we can make a qualitative comparison. Thus, for successful simulations, we examine the transient crater size at 0.1~s relative to the predicted crater size. The results are plotted as empty symbols in Figure~\ref{fig_crater}.

Unsurprisingly, the transient crater sizes at 0.1~s are all smaller than the craters predicted from the final time steps of our simulations, and thus the truth simulation lines up with the transient crater sizes quite well (Figure~\ref{fig_crater}a). While there is a relationship between the transient crater size and $Y_{\text{s}0}$, it is much weaker than the relationship between predicted crater size and $Y_{\text{s}0}$. In fact, for simulations with asteroid masses within 10\% of the truth simulation, all the transient crater widths are within 15\% of the truth simulation crater width. Thus, our results suggest that the early transient crater size does not significantly improve constraints on the material properties of the asteroid, but the final crater appears diagnostic in conjunction with $\Delta V$.

\subsection{Deflected or disrupted?}

When post-impact asteroid velocities are as large as the ones discussed in this study, the likelihood of disruption rather than deflection is an additional important caveat. An asteroid velocity change of 10\% of the escape velocity is commonly considered a safe threshold for deflection \citep[e.g.,][]{dearborn2015}, but the $\Delta V$ of the truth simulation is nearly 40\% of the escape velocity estimated for Dimorphos (5 cm/s). The more porous asteroids considered in this study would have even lower escape velocities, making the truth simulation $\Delta V$ as high as 73\% of escape velocity for the most porous asteroids. The geometry of the inverse test, with the impact vector perpendicular to the long axis of an asteroid with a large aspect ratio, also increases the probability of the asteroid breaking into distinct pieces.

Within the set of successful simulations, the possibility of disruption is particularly stark for the largest craters, where the diameter of the crater rim approaches the length of the minor axis of the model asteroid. In these cases, a large proportion of the asteroid (up to more than 50\%) is damaged, with only minor regions of undamaged material a significant distance from the impact site. When the strength of damaged material is low, the large regions of damaged material are unlikely to remain coherent long after impact because of minor variations in velocity across the asteroid body. 

While a detailed discussion of disruption is beyond the scope of this study, it is informative to examine the successful simulations for the probability of disruption. We assume that an asteroid that is $>$25\% damaged at the final analyzed time step would be at risk of disruption as the damaged regions in these simulations transect the entire diameter of the asteroid along the impact axis. Among the successful simulations, six meet this criterion for disruption risk. This set includes the four solutions with large craters and low values of $Y_{\text{s}0}$. However, similarly large craters that occur at larger values of $Y_{\text{s}0}$ have much smaller regions of damaged material. In fact, there is no single parameter that determines the extent of the damaged region. For instance, high porosities are generally associated with smaller damaged regions in our simulations, and no simulation in our disruption-risk set with a $\Delta V$ close to the truth simulation has a porosity greater than 0.4. However, there isn't a direct correlation between porosity and damage extent. Similarly, while the set of successful simulations at risk of disruption includes the four simulations with the lowest values of $Y_{\text{s}0}$, two simulations with higher values of $Y_{\text{s}0}$, including one with $Y_{\text{s}0}$~=~107~MPa, are also at high risk of disruption. Additional work is necessary to establish disruption metrics, including the effects of asteroid geometry and material properties on disruption probability.

In terms of the DART mission, however, we should emphasize again that the scenario presented in this test involves a significantly larger deflection than we expect in the actual DART experiment. This discrepancy largely results from the fact that the asteroid shape model used here is much smaller than the best estimates for Dimorphos \citep{naidu2020}. Based on the best available models for Dimorphos, we do not expect disruption to occur in the DART impact, though recent models of DART-like impacts into extremely low-strength targets indicate the possibility of global resurfacing \citep{raducan2022b}.

\section{Implications for DART and planetary defense}

The results of this inverse test demonstrate the vitality of knowing as much as possible about the asteroid target prior to attempting a kinetic deflection, particularly the asteroid’s mass, porosity, strength, and elastic properties. The results of this study describe a wide breadth of possible asteroid responses when the target volume, target shape, target material, impactor mass, impactor strength, and impactor velocity are defined. The velocity changes for our asteroid simulations, modeled across the inverse test as SiO$_2$, vary from $-0.89$~cm/s to $-4.6$~cm/s depending on the properties ascribed to the asteroid. Even when the mass of the target asteroid is constrained to $\pm$10\% of the truth simulation asteroid mass, the range of velocity changes is still $-0.89$~cm/s to $-3.3$~cm/s. Similarly, for successful simulations, with asteroid velocities close to the target $\Delta V$, there are a wide range of properties, with strengths ranging from a few kPa to more than more than 100~MPa and porosity ranging from 0.11 to 0.60.

Our inverse test results also demonstrate the need for robust extrapolations for $\Delta V$ and $\beta$, or, at the very least, robust metrics for choosing final values of $\Delta V$ and $\beta$. In the DART impact, $\Delta V$ will be measured in the weeks following impact, and for theoretical planetary defense missions, the efficacy of a kinetic deflection may be determined over months to years. However, running hydrocodes out to even a few seconds after impact can take days to weeks depending on the material parameters, problem scale, simulation resolution, and available resources. The late-time behavior of a body post-impact (e.g., a velocity change due to slow-moving ejecta) may be critical when planning deflection missions.

Additional information will be available from the DART mission that was not used in the inverse test. In addition to the mass estimate and crater observations by Hera discussed above, LICIACube will follow a few minutes behind the DART spacecraft to image the ejecta cone \citep{dotto2021}. We did not utilize information about the ejecta cone in the inverse test in order to model what can be learned from the minimum DART mission requirements, but the size and morphology of the ejecta cone may help constrain many properties. In particular, the ejecta cone images may easily rule out extreme material properties, when either a minuscule or massive amount of ejecta is produced. In addition, the ejecta behavior may provide information about the shallow subsurface makeup of Dimorphos. For instance, a large boulder directly below the impact site would likely decrease the amount of ejecta produced.

The general results of this study, e.g., the trends identified between the deflection magnitude and asteroid material parameters as well as the degeneracy associated with predicting material properties from a single observation, will be applicable to the DART impact regardless of the nature of the target body, Dimorphos. However, the precise numbers used and generated in this study are not expected in the DART impact. For instance, previous work shows that using a sphere as an impactor leads to a larger $\Delta V$ than would be expected from the DART spacecraft, with a sphere overestimating $\Delta V$ by up to 25\% \citep{owen2022,raducan2022c}. Another over-simplification in the models presented here is modeling the internal structure of the asteroid as homogeneous: rubble-pile structures can have important effects on ejecta production \citep[e.g.,][]{graninger2021,stickle2017}. Finally, the Itokawa shape model in this study was small relative to the prediction for Dimorphos, leading to larger deflections than predicted for DART. Thus, the expected risk for disruption or destabilization of the orbit of Dimorphos around Didymos is low.

While the miniature Itokawa simulated for the inverse test is likely small compared to Dimorphos, it is large enough to potentially require mitigation efforts were it on an impact trajectory with Earth. Depending on the albedo of this theoretical Dimorphos-sized hazardous object, the warning time in such a scenario could be short, requiring a deflection of the magnitude used for this inverse test and thus a non-negligible risk of disruption. The results of this study suggest that, in addition to understanding the effects of asteroid material properties, the role of asteroid geometry and impact location may be critical in analyzing disruption risk.

At the moment, significant uncertainties remain in the possible assemblages of material properties for hazardous asteroids, and, apart from computational studies, we know little about how these properties will affect kinetic deflection outcomes. DART will provide an essential first glimpse at this technology's potential, but its applicability across the diverse population of Near-Earth Asteroids will remain uncertain without future reconnaissance and mitigation demonstration missions. As emphasized in the 2023-2032 Planetary Science and Astrobiology Decadal Survey (\citeyear{decadal}), the ability to demonstrate rapid reconnaissance of asteroids will provide an important link between discovery and design of optimal mitigation missions. The computational study presented here provides further support for the importance of pre-mitigation characterization data on specific asteroid targets, while reinforcing the need to characterize the geotechnical properties of more Near-Earth Asteroids in general. The more information we have on likely impact scenarios prior to a potential deflection, the more likely a mission to defend the planet will be successful.

\begin{acknowledgments}
This work was supported by the DART mission, NASA Contract No. 80MSFC20D0004. 
Lawrence Livermore National Laboratory is operated by Lawrence Livermore National Security, LLC, for the U.S. Department of Energy, National Nuclear Security Administration under Contract DE-AC52-07NA27344. LLNL-JRNL-837536.
This work was also supported in part by the Advanced Simulation and Computing (ASC)--Verification and Validation (V\&V) program and ASC--Integrated Codes (IC) program at Los Alamos National Laboratory and a Chick Keller Postdoctoral Fellowship in Earth and Space Science from the Center for Space and Earth Science (CSES) at Loa Alamos National Laboratory. Los Alamos National Laboratory, an affirmative action/equal opportunity employer, is operated by Triad National Security, LLC, for the National Nuclear Security Administration of the U.S. Department of Energy under contract 89233218NCA000001. LA-UR-22-28492
\end{acknowledgments}

\clearpage

\appendix
\counterwithin{figure}{section}
\counterwithin{table}{section}

\section{The truth simulation}

\begin{figure*}[!b]
\fig{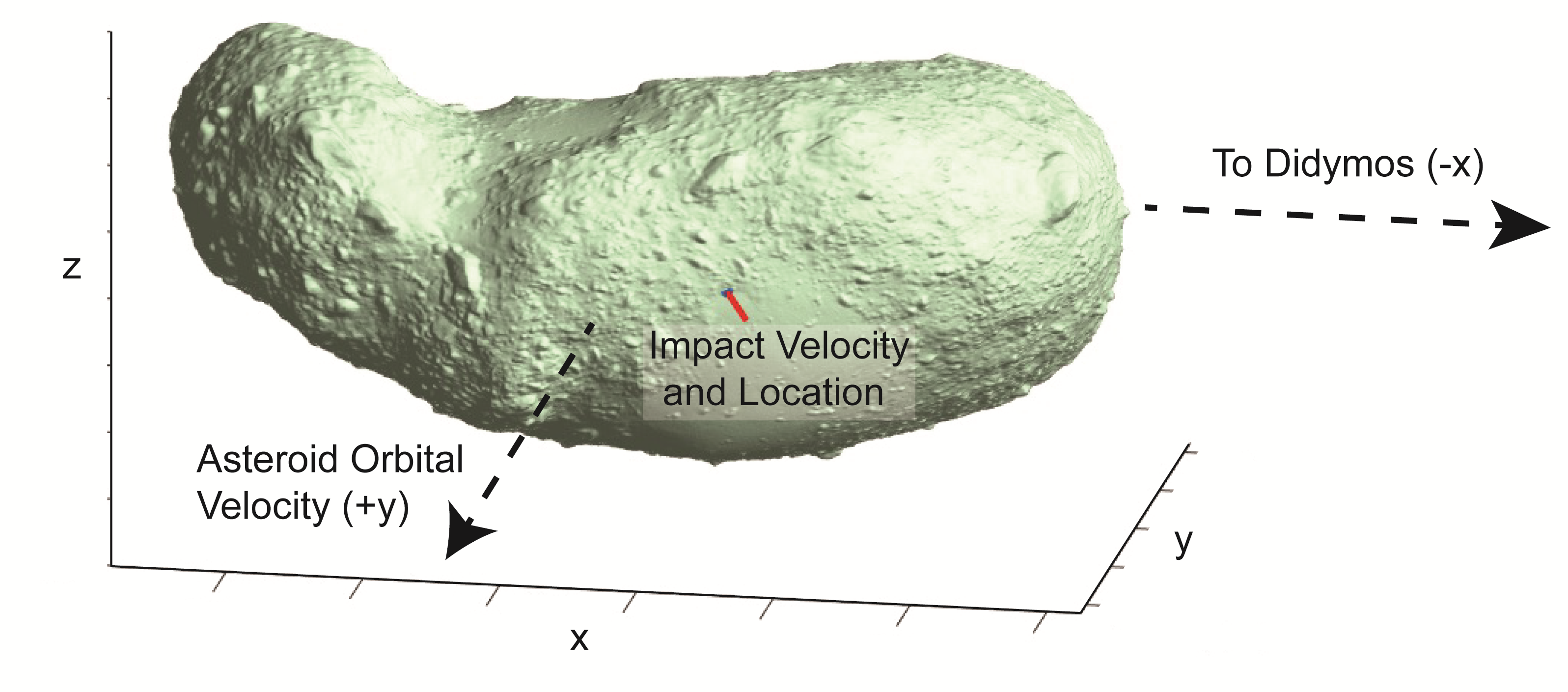}{0.75\textwidth}{}
\caption{Initial reference frame for this study, showing the shrunken Itokawa shape model and indicating the impact velocity vector (in red), the impact location (in blue), the vector toward the Didymos primary asteroid, and the vector of the asteroid's orbital velocity.}
\label{fig_BF}
\end{figure*}

The shape model used in this study to represent Dimorphos was a miniature Itokawa, with a total volume of 276,386~m$^3$ and a resolution of 2.31~m per facet. The model was provided in a reference frame such that the positive X direction was pointed away from the Didymos primary asteroid and the positive Y direction was parallel to the orbital velocity of Dimorphos (Figure \ref{fig_BF}). In this reference frame, the impact location was at (5.57$\times$10$^{-5}$,~0.029025,~$-$0.01614)~km, placing the impact location on a crater wall. The spacecraft velocity at impact was ($-$0.012,~$-$6.314,~3.511)~km/s. For the Spheral simulations, we rotated the simulation setup such that the spacecraft velocity was oriented in the negative Z direction for convenience; the Spheral reference frame is shown in Figure \ref{fig_visit}.

\subsection{CTH}

CTH is a two-step Eulerian shock physics hydrocode developed and maintained by Sandia National Laboratories \citep{mcglaun1990}. The CTH truth simulation uses a SESAME \citep{lyon1992} aluminum EOS to describe the impactor and a SESAME basalt EOS to describe the asteroid. The impactor has a radius of 36.51~cm, a mass of 612~kg, a von Mises strength of 275~MPa, and a tensile fracture strength of 310~MPa. The behavior of the asteroid is represented using a von Mises strength model with a strength of 10~MPa, a tensile fracture strength of 10~MPa, and a Johnson-Cook damage model with failure at a strain of 0.05. Porosity is modeled using a pressure-alpha model using parameters from \citet{jutzi2008}. The resolution of the simulation is set to 3 cells per projectile radius (cppr) on a flat mesh.

\subsection{FLAG}

FLAG (Free LAGrange) is a multiphysics Arbitrary Lagrangian-Eulerian code maintained by Los Alamos National Laboratory \citep{burton1992, burton1994a, burton1994b}. FLAG has been verified and validated for impact crater and high-velocity impacts and has been used to study planetary science applications \citep{caldwell2018, caldwell2019, caldwell2020, caldwell2021}. The FLAG truth simulation uses a Mie-Gr\"{u}neisen EOS for both the aluminum impactor as well as the basalt asteroid. Porosity is described with a pressure-alpha model, artificial viscosity with a von Neumann Richtmeyer formulation with $q2$~=~1.3 and $q1$~=~0.3, and damage with a Johnson-Cook model with $d1$~=~0.05 \citep{johnson1985}. A Tipton closure model is used to handle mixed-material zones. The resolution of the simulation ranged from 7.3~cm (5~cppr) at the impact site to 1~m (0.37~cppr) away from the impact.

\section{Spheral material models and metrics: Additional information}

\subsection{Material models}

We use a 5th order B-spline kernel and our adaptive smoothing length algorithm maintains roughly 4 radial neighbors spanning the extent of the kernel \citep{owen2010}. This equates to roughly 268 neighbors in three dimensions. We use Monaghan-Gingold \citep{monaghan1983} artificial viscosity with linear coefficient, $Cl$=1.5, and quadratic coefficient, $Cq$=0.75. Strength and shear modulus are modified by the porosity as $n = n_{0}(1-\phi)$, with $n$ being either the strength or shear modulus of intact or damaged material. The coefficient of internal friction is set to 1.2 for intact material and 0.6 for damaged material.

In our strain-porosity model, compaction is treated as elastic up to a strain of $–1.88 \times 10^{-4}$, beyond which the compaction is modeled as exponential with exponential compaction factor $\kappa$=0.9. In our damage model, our flaw distribution is governed by the proportional and exponential Weibull constants, $k$ and $m$ respectively. Flaws become active and propagate damaging the material once their activation strain is exceeded. In this paper, we set $k$ and $m$ equal to $5.00\times10^{24}$ and 9.0, respectively. The activation strain is calculated using Spheral's ``Pseudo Plastic Strain" option. This involves using the time-derivative of deviatoric stress and the shear modulus to integrate the activation strain in time. Additional details regarding the damage model and the parameters used can be found in \citet{owen2022}.

The impactor is modeled as a solid aluminum sphere with no porosity and a radius of 37.8 cm (calculated from the density of solid aluminum at 2.7 g/cm$^3$ and an impactor mass of 612 kg). Most simulations in this paper use a LEOS (Livermore Equation of State) aluminum EOS to describe the impactor and a LEOS SiO$_2$ EOS to describe the asteroid \citep{fritsch2016}. Simulations testing EOS choice use either a LEOS aluminum EOS or a Tillotson aluminum EOS to describe the impactor (the choice of impactor EOS had little effect on the impact simulation). Asteroid EOS and material combinations for these simulations are ANEOS SiO$_2$ \citep{thompson1990}, Tillotson basalt, Tillotson granite, and Tillotson pumice \citep{tillotson1962}. Parameters used in the Tillotson EOS for different phases can be found in Table \ref{atab_till}.

\begin{deluxetable}{llll}
\tablecaption{Parameters used in Tillotson EOS}
\tablehead{
\multirow{2}{*}{Parameter} & \multicolumn{3}{c}{Material} \\
\cline{2-4}
 & Basalt & Granite & Pumice
}
\startdata
$\rho_0$ [g/cm$^3$] & 2.700 & 2.680 & 2.327 \\
a & 0.5 & 0.5 & 0.5 \\
b & 1.5 & 1.3 & 1.5 \\
A [dyne/cm$^2$] & 2.67e11 & 1.80e11 & 2.67e11 \\
B [dyne/cm$^2$] & 2.67e11 & 1.80e11 & 2.67e11 \\
$\alpha$ & 5.0 & 5.0 & 5.0 \\
$\beta_{\text{EOS}}$ & 5.0 & 5.0 & 5.0 \\
$\varepsilon_0$ [erg/g] & 4.87e12 & 1.60e11 & 4.87e12 \\
$\varepsilon_{\text{liquid}}$ [erg/g] & 4.72e10 & 3.50e10 & 4.72e10 \\
$\varepsilon_{\text{vapor}}$ [erg/g] & 1.82e11 & 1.80e11 & 1.82e11 \\
\enddata
\tablecomments{$\beta_{\text{EOS}}$ is a parameter in the Tillotson EOS \citep{tillotson1962} and is entirely separate from the momentum enhancement factor $\beta$ discussed in this paper.}
\label{atab_till}
\end{deluxetable}
\vspace{-2.5em}

\subsection{Ejecta definition} \label{sec:ejecta}

\begin{figure*}
\gridline{\fig{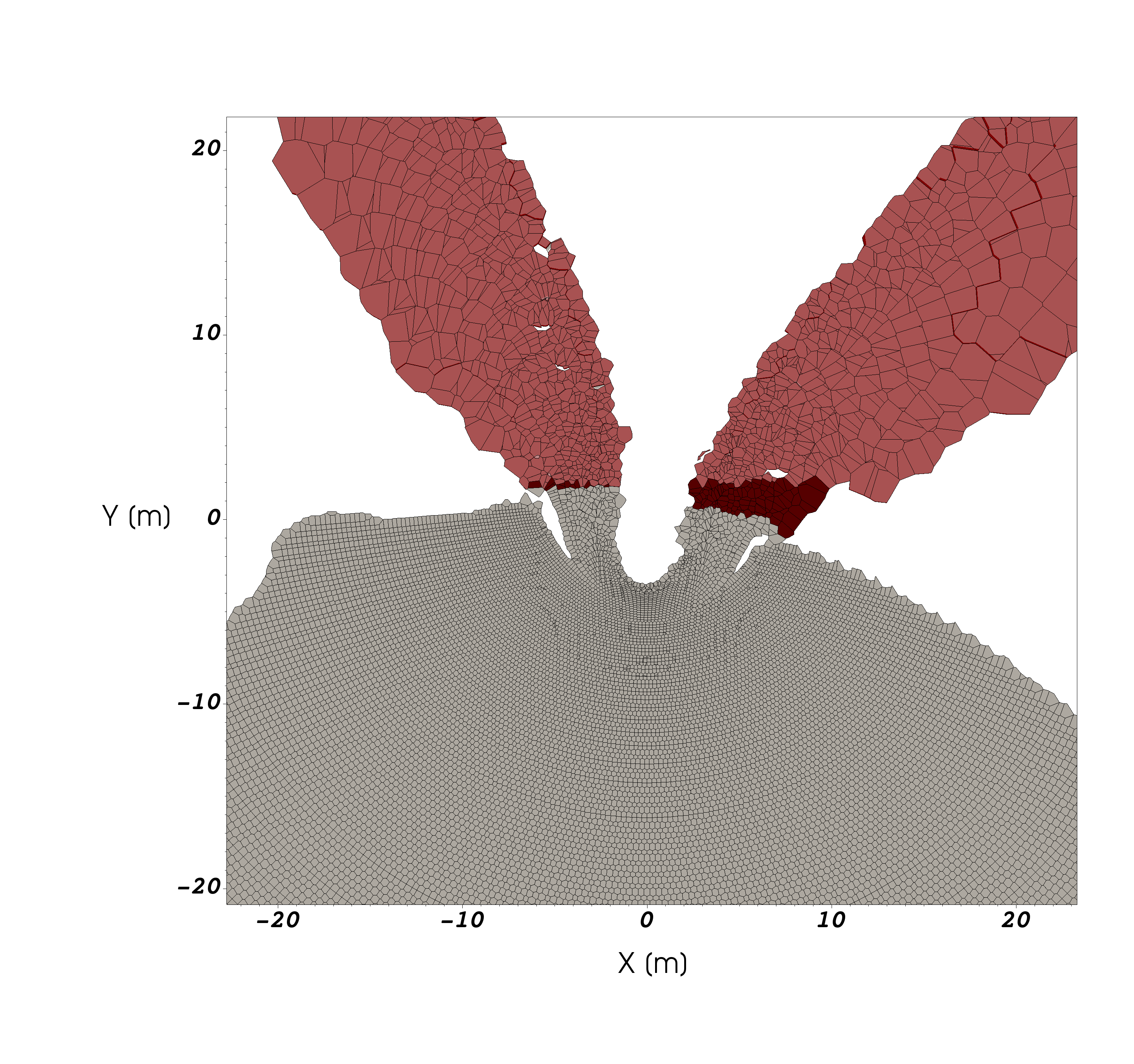}{0.47\textwidth}{(a)}
          \fig{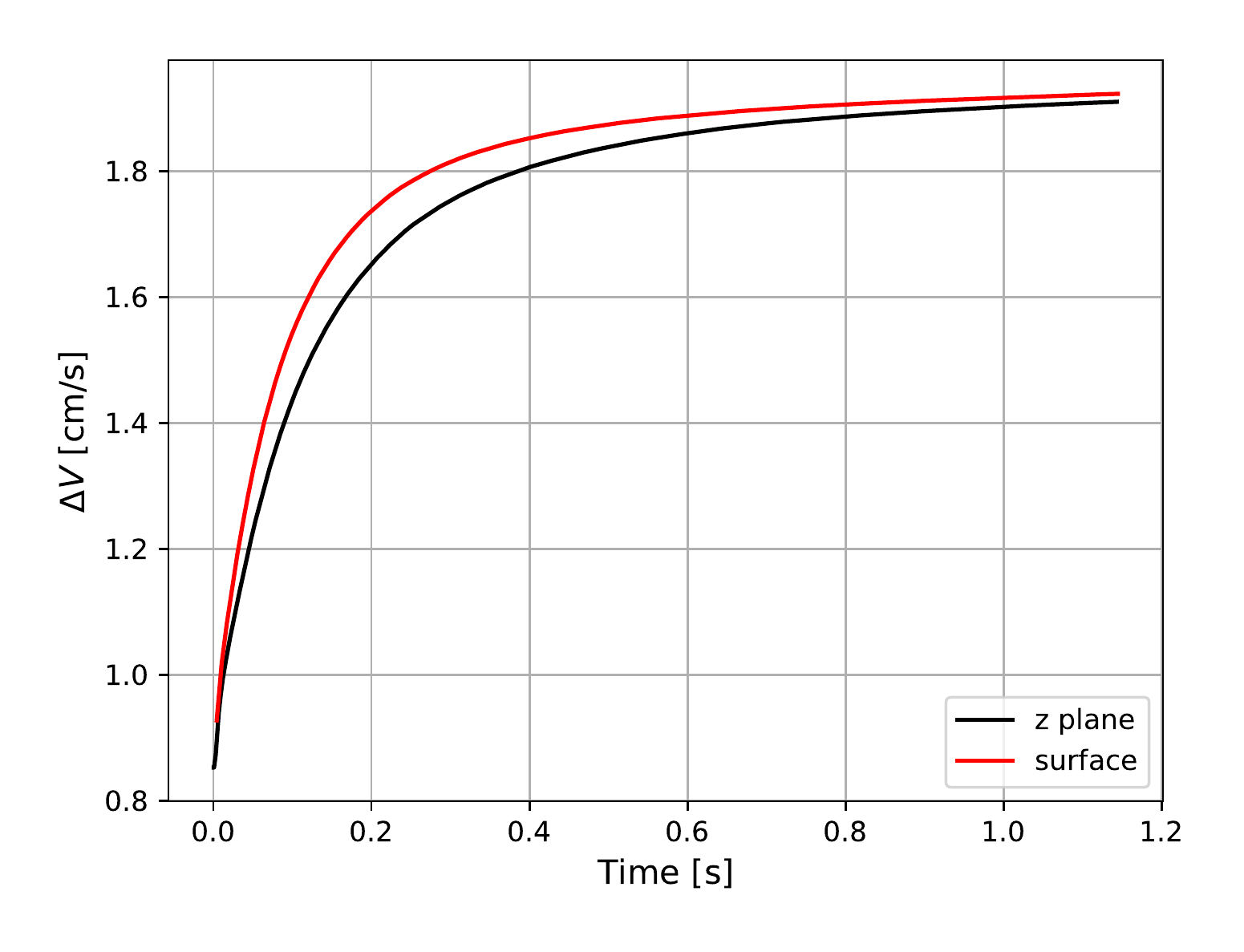}{0.54\textwidth}{(b)}
          }
\caption{(a) Example simulation exhibiting the difference in ejecta definition. Material in grey is considered part of the asteroid and material in light red is considered ejecta in both definitions. The material in dark red is considered ejecta only by our definition using the asteroid surface. (b) $\Delta V$ convergence with the old z-plane definition (black) and the new surface definition (red).}
\label{fig_ejecta}
\end{figure*}

Many hydrocodes define ejecta using two filters: velocity and spatial location. The first filter checks that the velocity of the material is greater than the escape velocity for the target being struck. The second filter is a plane above the asteroid surface with a normal parallel to the spacecraft velocity vector. Material above the plane moving faster than the escape velocity is considered ejecta and contributes to the ejecta momentum (Figure \ref{fig_ejecta}a).

We used a planar filter for ejecta at the outset of this study. However, the geometry of the shrunken Itokawa asteroid meant that a portion of the early ejecta traveled at a shallow angle relative to the spacecraft velocity vector. This trajectory meant this material was not accounted for using the typical ejecta definition (Figure \ref{fig_ejecta}a). For our simulations, therefore, we incorporated a filter parallel to the asteroid surface, recalculating the results of previous simulations. Material moving faster than the escape velocity and located at least 1~m above the original asteroid surface was labeled as ejecta. In addition to capturing fast-moving low-angle ejecta, this ejecta definition had the added benefit of converging to a stable $\Delta V$ and $\beta$ more quickly (Figure \ref{fig_ejecta}b). We expect the magnitude of this benefit to be directly related to the geometry of the asteroid and the angle of impact relative to the asteroid surface orientation.

\subsection{Crater definition}

\begin{figure*}
\fig{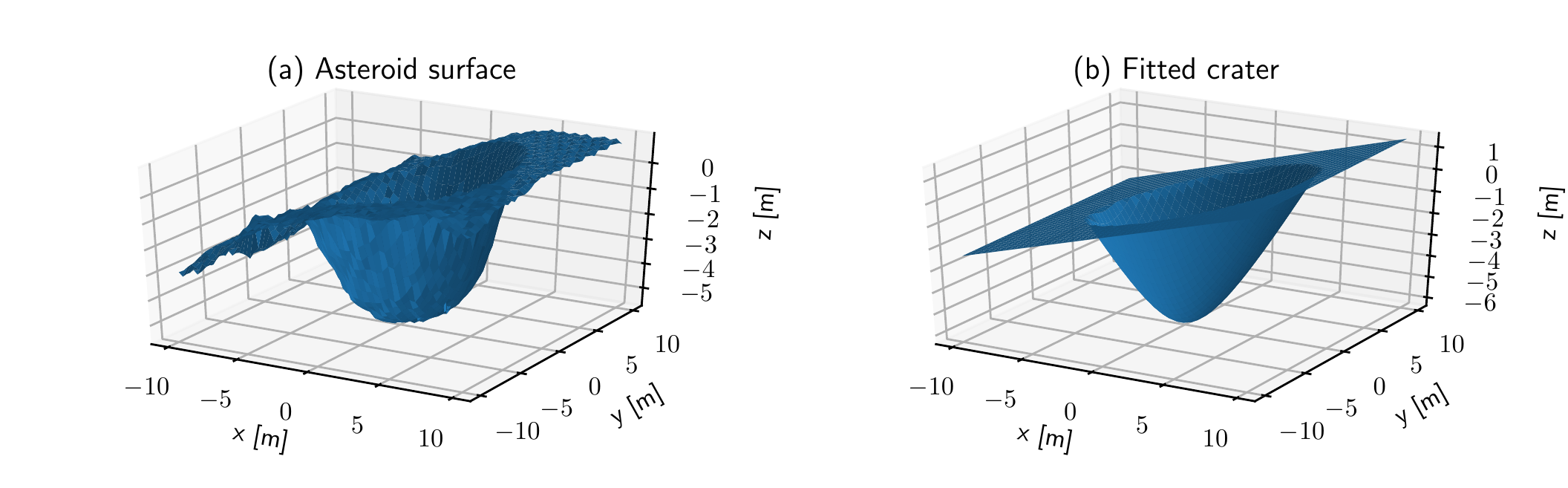}{0.8\textwidth}{}
\caption{Example of crater fit for simulation 11.09. (a) Original surface of critical density. (b) Plane/hyperboloid fit to the surface.}
\label{fig_craterFit}
\end{figure*}

Crater morphology was described for simulations that produced the correct $\Delta V$. Small craters ($<$20~m in diameter) were fit using a hyperboloid to describe the crater itself and a plane to describe the asteroid surface. The crater depth was defined as the distance between the plane and the hyperboloid below the plane. Crater diameter was defined as the average of the two axes of the ellipse formed by the intersection of the plane and the hyperboloid \citep{klein2013}. While the craters in the simulations were not always hyperbolic in morphology, the crater depth and diameter fit using this method were accurate for many different crater geometries (Figure~\ref{fig_craterFit}). For large craters with diameters greater than 20~m, the volume of excavated material was large enough to invalidate the planar assumption for the asteroid surface. In these cases, crater size was measured manually from two perpendicular cross-sections.


\section{Using machine learning to examine the search space}

For the first 138 simulations, $Y_{\text{s}0}$ and $G_{\text{s0}}$ were set to cover ranges more equivalent to intact or partially fractured rock (Table~\ref{atab_search}). For next 120 simulations, however, we focused on weaker materials more similar to sand or clay. The ranges for $P_{\text{min}}$, $P_{\text{d,min}}$, and $\phi$ did not vary across these first 258 simulations. The final 80 simulations included information on asteroid mass, thus constraining porosity. Other variables in these final simulations were allowed to vary across the full ranges explored in this study.

We set up a machine learning decision tree algorithm to direct the parameter choices tested in this study, using the seven material parameters as inputs and $\Delta V$ as the output. The algorithm was initially seeded with a set of 40 simulations with randomized parameter combinations (Scan 1). Ten successive scans of 3--16 simulations each were run for the ``intact rock" parameters, nine scans of 8–-16 simulations each were run for ``fractured rock" parameters, and five scans of 16 simulations each were run for the mass-constrained parameters.

For each scan, the algorithm generated between 5,000 and 10,000 possible parameter combinations that could potentially produce the correct $\Delta V$ using a decision tree with a tree depth of 34. We then selected the simulations for that scan using Mitchell's Best Sampling algorithm \citep{mitchell1991}. This selection algorithm uses a k-d tree to find the simulations that are the furthest apart in the seven-dimensional input space. Once the simulations for each scan were run and we were satisfied with the extrapolations for $\Delta V$, the results were added to the training set for the decision tree algorithm before choosing material parameter combinations for the next scan. 

We note that we changed the ejecta definition used to calculate $\Delta V$ after Scan 17. Details regarding the two ejecta definitions are described in Section \ref{sec:ejecta}. All reported values for $\Delta V$ and $\beta$ are calculated using the new ejecta definition, but the decision tree algorithm we utilized likely would not reproduce the same material parameter sets tested in this study.

\begin{deluxetable}{llll}
\tablecaption{Variable ranges searched for simulations}
\tablehead{
Variable & Strong intact rock (n = 138) & Weak fractured rock (n = 120) & Mass-constrained (n = 80)
}
\startdata
$Y_{\text{s}0}$ [MPa] & $10^{0} - 10^{2.2}$ (linear) & $10^{-3} - 10^{0}$ (logarithmic) & $10^{-3} - 10^{2.2}$ (logarithmic)\\
$Y_{\text{d}0}$ [MPa] & $10^{-4} - 10^{0}$ (linear) & $10^{-5} - 10^{-1}$ (logarithmic) & $10^{-5} - 10^{0}$ (logarithmic)\\
$G_{\text{s0}}$ [GPa] & $10^{-1} - 10^{2}$ (linear) & $10^{-2} - 10^{1}$ (logarithmic) & $10^{-2} - 10^{2}$ (logarithmic)\\
$G_{\text{d0}}$ [GPa] & $10^{-3} - 10^{0}$ (linear) & $10^{-4} - 10^{0}$ (logarithmic) & $10^{-4} - 10^{3}$ (logarithmic)\\
$-P_{\text{min}}$ [GPa] & $10^{-8} - 10^{-1}$ (logarithmic) & $10^{-8} - 10^{-1}$ (logarithmic) & $10^{-8} - 10^{-1}$ (logarithmic)\\
$-P_{\text{d,min}}$ [GPa] & $10^{-10} - 10^{-4}$ (logarithmic) & $10^{-10} - 10^{-4}$ (logarithmic) & $10^{-10} - 10^{-4}$ (logarithmic)\\
$\phi$ & $0.05 - 0.70$ (linear) & $0.05 - 0.70$ (linear) & $0.10 - 0.26$ (linear) \\
\enddata
\tablecomments{Whether the search was conducted in linear or logarithmic space is noted for each variable in each simulation set. $P_{\text{min}}$ and $P_{\text{d,min}}$ are tensile pressures.}
\label{atab_search}
\end{deluxetable}
\vspace{-2em}

\section{Extrapolations to large times}

\begin{figure*}[!h]
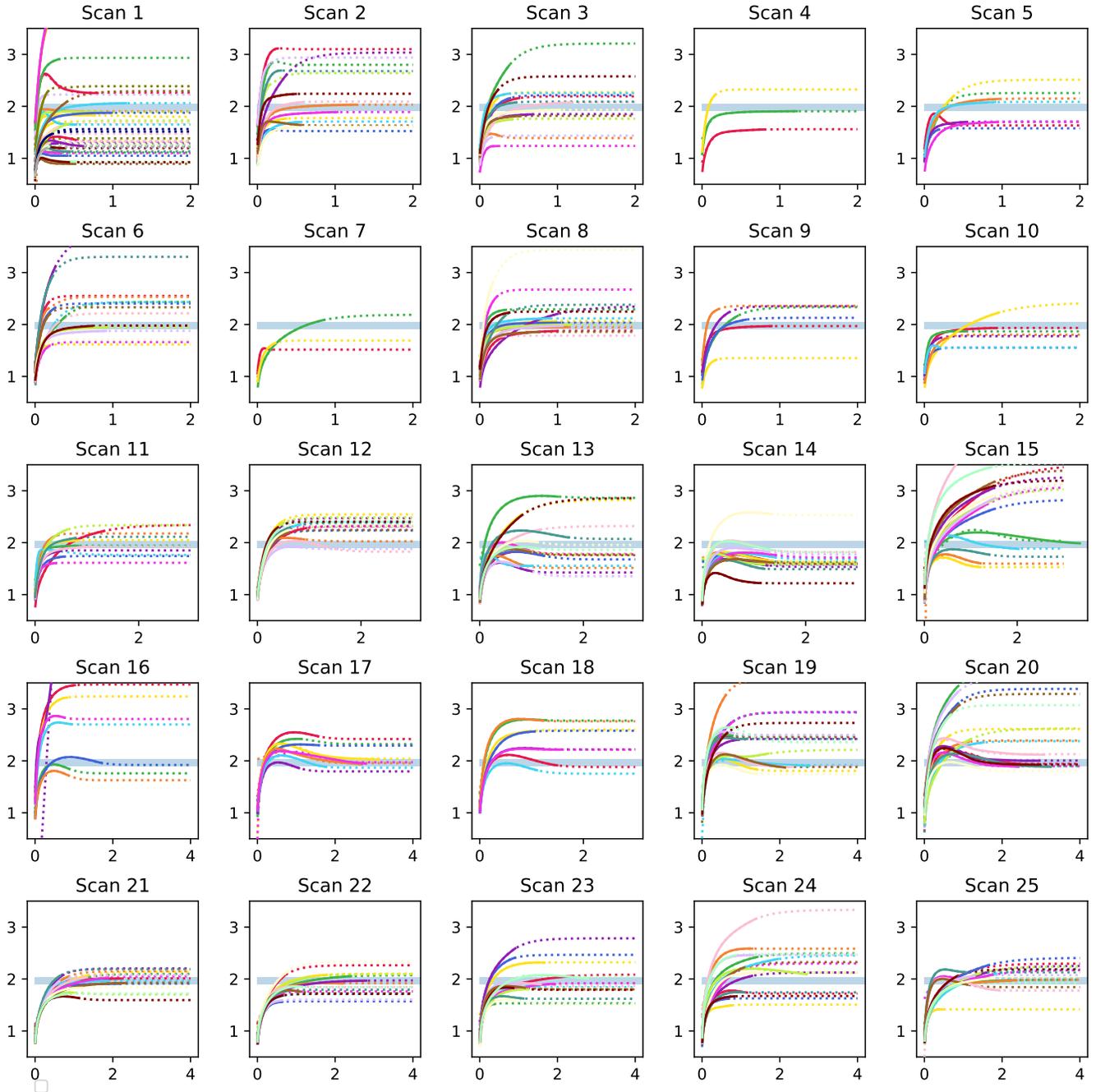

\fig{allExtrap.pdf}{\textwidth}{}
\caption{Extrapolations for $\Delta V$ for all simulations, with time in seconds on the X-axis and the magnitude of $\Delta V$ in centimeters per second on the Y-axis. Solid lines show the data from each simulation, and dotted lines in the same color show the fit of the extrapolation function to those data. The blue bar at the center of each plot shows the target $\Delta V$ of 1.972~$\pm$~0.07~cm/s.}
\label{fig_extrap}
\end{figure*}

The large parameter space examined in this study necessitates running many simulations to adequately sample the multi-dimensional space. Thus, due to computational constraints, simulations could not always be run until the system equilibrated. For the purposes of the machine learning algorithm, we fit curves of $\Delta V$ versus time with an exponential decay model. A Gaussian was added to phenomenologically describe the local maximum observed at early times. The curve was fit as:
\begin{equation}
\Delta V = C_1 \times [1 - \text{exp}(-C_2(t+C_3))] + C_4 \times \text{exp} \left [ - \frac{(t-C_5)^2}{2C_6} \right ]
\end{equation}
where $t$ is time and $C_1$ through $C_6$ are fitting constants. $\Delta V$ at $t_{\infty}$ is equal to $C_1$. Figure \ref{fig_extrap} depicts the fits to all of the simulations run for the inverse test, demonstrating how we extracted extrapolated values for $\Delta V$ and $\beta$ from simulations that do not run very long. For simulations that significantly undershot or overshot the target $\Delta V$ (the blue bar), in particular, these extrapolations gave us more realistic velocity magnitudes to input into the machine learning algorithm.

The extrapolations we've used here are limited in scope and likely predict convergence at earlier times than expected for true system equilibrium. Even simulations run out to times of $>$3~s do not capture the slow-moving ejecta that may leave the system at late times. The escape velocity of Dimorphos is estimated to be 5~cm/s. Since our ejecta definition required material to be at least 1~m above the asteroid surface before it would be calculated as ejecta, the slowest-moving proto-ejecta particles would take at least 20~s to pass this boundary. This would take weeks of computation time on hundreds of processors and would not be usable for the quick turnarounds desired for planetary defense missions, including DART.

That said, while the velocity of this ejecta will clearly be much slower than the ejecta analyzed, significant mass may still escape the asteroid and contribute to a momentum change. For instance, in one simulation of a weak asteroid (22-05), the projected final crater was $\sim$40~m wide and excavated $\sim$15~m down, with $\sim$6\% of the asteroid's mass predicted to eventually escape. This is dramatically different from the $\sim$0.46\% of mass that had escaped to this point, despite the comparatively flat relationship between $\Delta V$ and time. Thus, the overall magnitude of the contribution of this late-stage ejecta must be analyzed in future studies.

\section{Extremes in behavior}

While the exponential decay model used to describe the late-time behavior of our simulations works well for most of our parameter combinations, this is not true for all simulations. Simulations of weak asteroid targets with significant ejecta coming from locations other than the impact site were especially poorly fit by the exponential decay model. The extra ejecta typically decreases $\Delta V$ and $\beta$ in the orbital direction with time, resulting in a lack of convergence with time. For these simulations, reported values for $\Delta V$ and $\beta$ are for the final time step (commonly $>$2~s), but they are expected to be lower.

\begin{figure*}
\fig{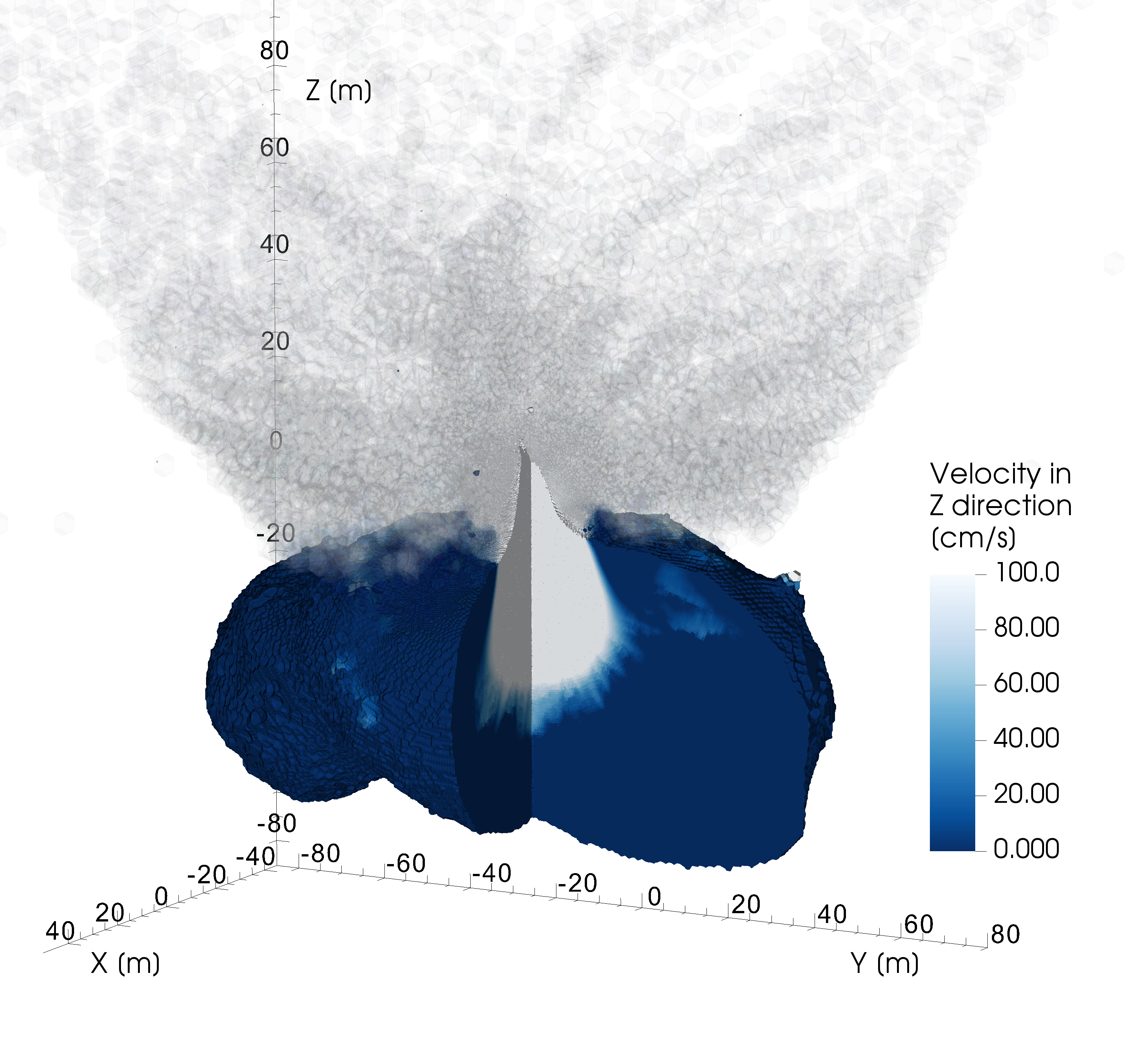}{0.6\textwidth}{}
\caption{Fluid-like behavior in simulation 16.05. Asteroid material is colored according to its velocity in the Z direction (with white material moving upward at or faster than 1~m/s). Material with a density below 0.5~g/cm$^3$ is shown in translucent black. One quarter of the simulation visualization is clipped out to show the velocity fields interior to the asteroid.}
\label{fig_fluid}
\end{figure*}

One simulation is an extreme outlier, with nearly fluid-like behavior (Figure~\ref{fig_fluid}). Many of the parameters for this simulation are well within the ranges describing successful simulations, except for the damaged shear modulus ($G_{\text{d0}}$~=~174~kPa). This modulus is even smaller than low-density aerogel \citep[e.g.,][]{scherer1995} and as such is unsuitable for describing a realistic asteroid response.

\clearpage
\section{Additional supplemental figures and results}

\begin{figure*}[!h]
\fig{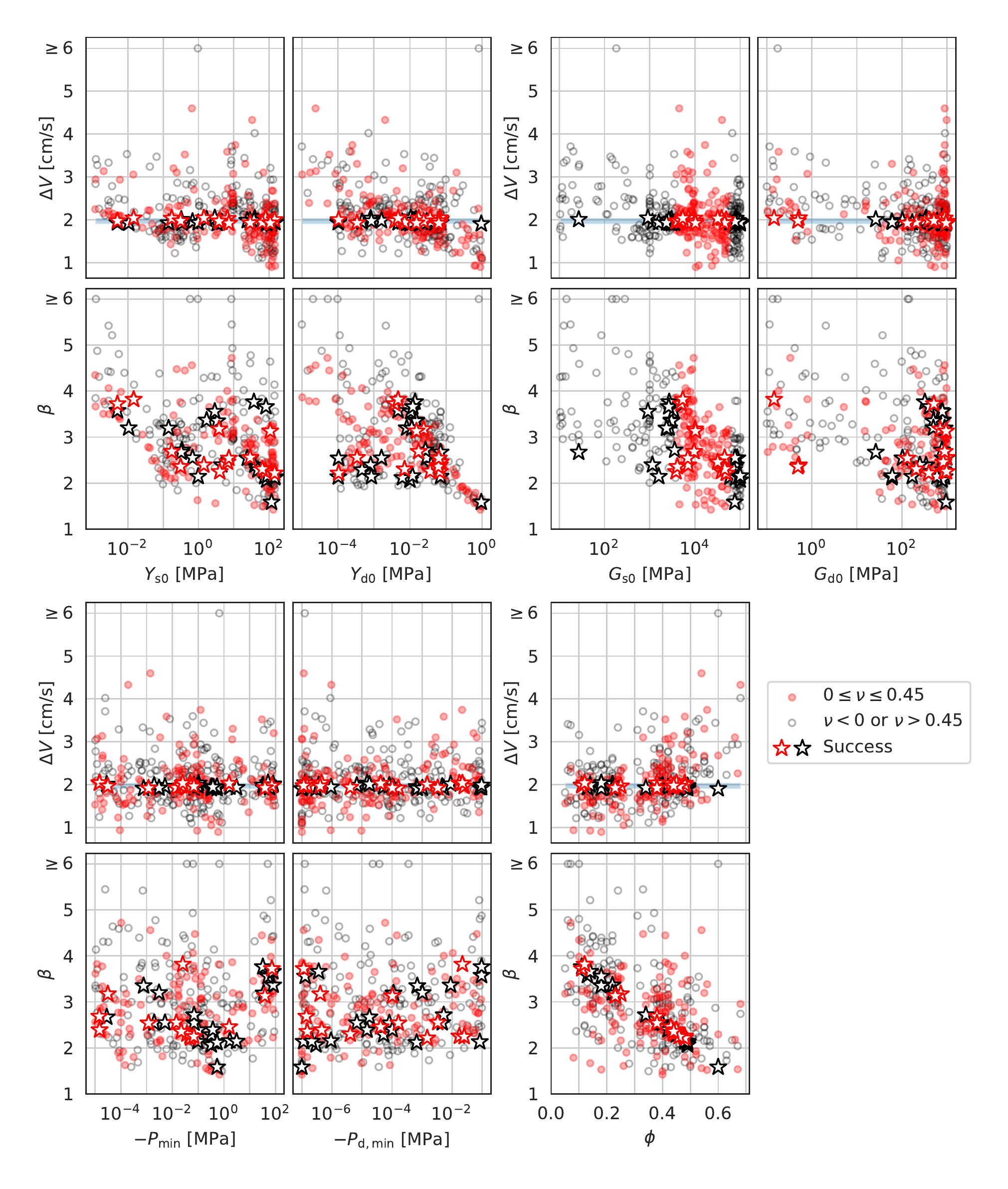}{0.85\textwidth}{}
\caption{Trends across all simulation results. This figure contains the same simulation results as Figure \ref{fig_weakstrong}, but points are colored in red for Poisson's ratios between 0 and 0.45 and black for lower or higher Poisson's ratios. Stars indicate simulations with extrapolated values of $\Delta V$ in the success range.}
\label{fig_poisson}
\end{figure*}

\begin{figure*}[!h]
\fig{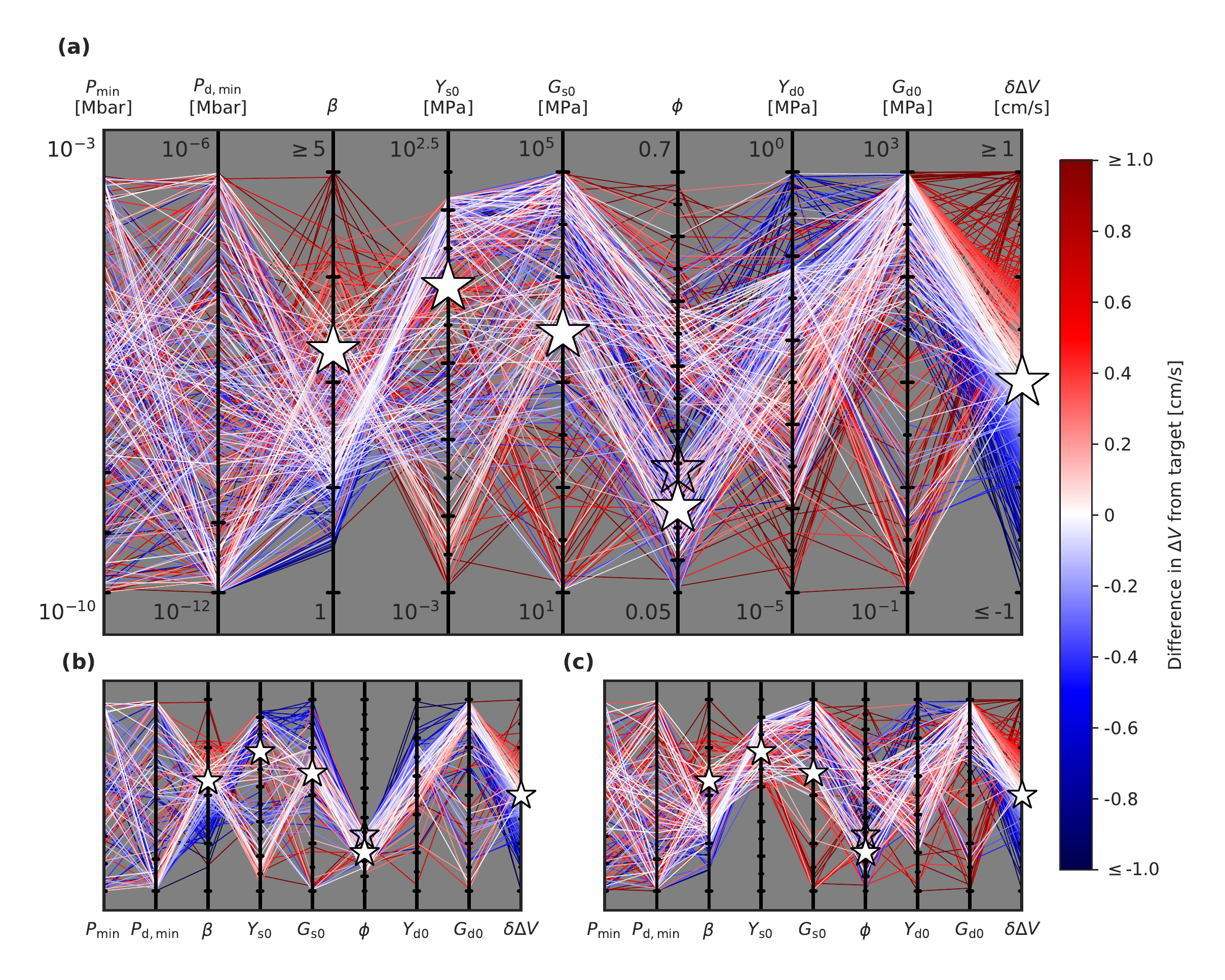}{\textwidth}{}
\caption{Parallel coordinate plots showing all input variables of the inverse simulations, as well as final $\Delta V$ and $\beta$. Simulation lines are colored by the difference between the simulation $\Delta V$ and the target $\Delta V$, with blue being slower than the truth simulation and red being faster than the truth simulation. White stars indicate the values of the truth simulation. For $\phi$, the open star indicates the original truth simulation value of 0.24 while the solid white star indicates the porosity required for an equivalently massive asteroid in our simulations. Plots are shown for (a) all simulations, (b) simulations with initial asteroid mass within 10\% of the truth simulation, and (c) simulations with $Y_{\text{s0}}$ within one order of magnitude of the truth simulation. Ticks and axis limits in (b) and (c) are the same as in (a).}
\label{fig_spaghettiall}
\end{figure*}

\begin{splitdeluxetable}{lllllllllllBllllllll}[!h]
\tablecaption{Inverse simulation inputs and results}
\tablehead{
Sim ID & Processors & Run time & Sim time & $Y_{\text{s0}}$ & $Y_{\text{d0}}$ & $G_{\text{s0}}$ & $G_{\text{d0}}$ & $\phi$ & $P_{\text{min}}$ & $P_{\text{d min}}$ 
& Sim ID & $\beta_{t=0.1}$ & $\beta_{t=\text{max}}$ & $\beta_{t=\infty}$ & $\Delta V_{t=0.1}$ & $\Delta V_{t=0.2}$ & $\Delta V_{t=\text{max}}$ & $\Delta V_{t=\infty}$ \\
& & [hr] & [s] & [MPa] & [MPa] & [MPa] & [MPa] & & [MPa] & [MPa] & & & & & [cm/s] & [cm/s] & [cm/s] & [cm/s]
}
\startdata
1.00 & 144 & 54.2	& 0.8038	& 81	& 0.785	& 4530	& 169	& 0.67	& $-$5.87e$-$4	& 0	&
1.00 & 1.7958	& 1.5308	&  1.5355	& $-$2.5920	& $-$2.5576	& $-$2.2568	& $-$2.2603 \\
1.01	& 144	& 38.5	& 0.2768	& 64	& 0.985	& 100000	& 380	& 0.35	& $-$8.38e$-$4	& 0	& 
1.01	& 1.5411	& 1.4957& 1.4951	& $-$1.1353	& $-$1.1170	& $-$1.1113	& $-$1.1112 \\
1.02	& 144	& 86.1	& 0.7703	& 113	& 0.375	& 52200	& 331	& 0.52	& $-$2.49e$-$4	& 0	& 
1.02	& 1.8247	& 1.8548	& 1.8571	& $-$1.7902	& $-$1.8212	& $-$1.8288	& $-$1.7982 \\
1.03	& 144	& 38.3	& 0.4156	& 119	& 0.99	& 28400	& 83	& 0.26	& $-$3.20e$-$4	& 0	& 
1.03	& 1.6932	& 1.6082	& 1.6092	& $-$1.0871	& $-$1.0557	& $-$1.0490	& $-$1.0489 \\
1.04	& 144	& 76.2	& 0.6123	& 117	& 0.952	& 76000	& 953	& 0.6	& $-$5.42e$-$4	& 0	& 
1.04	& 1.6247	& 1.5831	& 1.5835	& $-$1.9458	& $-$1.9361	& $-$1.9021	& $-$1.9035 \\
1.05	& 144	& 38.1	& 0.2900	& 118	& 0.716	& 88000	& 297	& 0.41	& $-$6.62e$-$4	& 0	& 
1.05	& 1.5656	& 1.5805	& 1.5818	& $-$1.2697	& $-$1.2788	& $-$1.2858	& $-$1.2869 \\
1.06	& 144	& 38.6	& 0.5289	& 38	& 0.859	& 16400	& 875	& 0.56	& $-$2.49e$-$4	& 0	& 
1.06	& 1.7659	& 1.5177	& 1.5199	& $-$1.8900	& $-$1.7730	& $-$1.6496	& $-$1.6507 \\
1.07	& 144	& 38.8	& 0.3272	& 50	& 0.976	& 64000	& 36	& 0.28	& $-$5.48e$-$5	& 0	& 
1.07	& 1.7115	& 1.6557	& 1.6572	& $-$1.1276	& $-$1.1065	& $-$1.0993	& $-$1.1000 \\
1.08	& 144	& 85.0	& 0.8513	& 132	& 0.286	& 40200	& 336	& 0.52	& $-$2.59e$-$4	& 0	& 
1.08	& 1.8740	& 1.9276	& 1.9279	& $-$1.8328	& $-$1.8841	& $-$1.8897	& $-$1.8900 \\
1.09	& 144	& 19.3	& 0.1623	& 108	& 0.58	& 70100	& 101	& 0.37	& $-$5.18e$-$4	& 0	& 
1.09	& 1.6870	& 1.6756	& 1.6639	& $-$1.2733	& 			& $-$1.2663	& $-$1.2572 \\
1.10	& 144	& 38.7	& 0.4465	& 131	& 0.897	& 22400	& 622	& 0.4	& $-$6.71e$-$4	& 0	& 
1.10	& 1.5759	& 1.4229	& 1.4213	& $-$1.2519	& $-$1.1772	& $-$1.1301	& $-$1.1312 \\
1.11	& 144	& 19.0	& 0.1516	& 143	& 0.602	& 94000	& 887	& 0.63	& $-$2.43e$-$5	& 0	& 
1.11	& 1.6997	& 1.7196	& 1.7269	& $-$2.1948	& 			& $-$2.2199	& $-$2.2256 \\
1.12	& 144	& 38.8	& 0.5089	& 118	& 0.896	& 10400	& 541	& 0.18	& $-$5.96e$-$4	& 0	& 
1.12	& 1.6795	& 1.5270	& 1.5303	& $-$0.9689	& $-$0.9117	& $-$0.8908	& $-$0.8935 \\
1.13	& 144	& 19.3	& 0.1743	& 103	& 0.686	& 58100	& 352	& 0.43	& $-$5.73e$-$4	& 0	& 
1.13	& 1.6232	& 1.5940	& 1.5833	& $-$1.3563	& 			& $-$1.3342	& $-$1.3242 \\
1.14	& 144	& 38.4	& 0.3587	& 69	& 0.871	& 46100	& 935	& 0.4	& $-$3.25e$-$4	& 0	& 
1.14	& 1.5851	& 1.4934	& 1.4949	& $-$1.2613	& $-$1.2154	& $-$1.1939	& $-$1.1951 \\
\enddata
\tablecomments{This table is published in its entirety in the machine-readable format. A portion is shown here for guidance regarding its form and content. The simulation ID column is repeated here for clarity.
The number of computer processors used as well as the computer run time required to reach the final simulation time are reported for each simulation. $\beta$ is reported at $t = 0.1$ s, $t=t_{\text{max}}$, and the extrapolated $t=t_{\infty}$. $\Delta V$ is reported at $t = 0.1$ s, $t = 0.2$ s, $t=t_{\text{max}}$, and the extrapolated $t=t_{\infty}$.}
\label{atab_alldata}
\end{splitdeluxetable}

\begin{deluxetable}{llcccccccc}
\tablecaption{Results of simulations comparing EOS/material choice}
\tablehead{
\multicolumn{2}{l}{Original sim ID:} & \multicolumn{2}{c}{1.04} & \multicolumn{2}{c}{9.00} & \multicolumn{2}{c}{11.09} & \multicolumn{2}{c}{22.05} \\
\cmidrule(lr){1-2} \cmidrule(lr){3-4} \cmidrule(lr){5-6} \cmidrule(lr){7-8} \cmidrule(lr){9-10}
\multirow{2}{*}{EOS} & \multirow{2}{*}{Material} & $\Delta V$ & $\beta$ & $\Delta V$ & $\beta$ & $\Delta V$ & $\beta$ & $\Delta V$ & $\beta$ \\
& & [cm/s] & & [cm/s] & & [cm/s] & & [cm/s] &
}
\startdata
LEOS & SiO$_2$ & $-$1.90 & 1.58 & $-$1.97 & 2.14 & $-$1.92 & 2.55 & $-$1.97 & 3.72 \\
ANEOS & SiO$_2$ & $-$1.94 & 1.63 & $-$1.98 & 2.18 & $-$1.80 & 2.40 & $-$1.93 & 3.65 \\
Tillotson & basalt & $-$2.01 & 1.73 & $-$2.06 & 2.30 & $-$2.07 & 2.82 & $-$2.07 & 4.00 \\
Tillotson & granite & $-$1.84 & 1.55 & $-$1.89 & 2.08 & $-$1.96 & 2.64 & $-$2.10 & 4.03 \\
Tillotson & pumice & --- & --- & $-$2.31 & 2.21 & $-$2.32 & 2.71 & $-$2.32 & 3.90 \\
\enddata
\tablecomments{Tillotson pumice not run for simulation 1.04. All values of $\Delta V$ and $\beta$ in this table are extrapolated to $t \sim \infty$, following the method used in the rest of this study.}
\label{atab_EOS}
\end{deluxetable}

\begin{deluxetable}{lllllll}
\tablecaption{Transient and final predicted crater dimensions for successful simulations}
\tablehead{
\multirow{2}{*}{Sim ID} & \multicolumn{3}{c}{Transient crater (0.1 s)} & \multicolumn{3}{c}{Final predicted crater} \\
\cmidrule(lr){2-4} \cmidrule(lr){5-7}
& x [cm] & y [cm] & z [cm] & x [cm] & y [cm] & z [cm]
}
\startdata
1.04\tablenotemark{a}	& 825	& 801	& 438	& 1148	& 1211	& 373 \\
2.04\tablenotemark{a}	& 754	& 719	& 438	& 1268	& 1338	& 508 \\
3.13\tablenotemark{a}	& 650	& 688	& 375	& 1124	& 1212	& 430 \\
4.01\tablenotemark{a}	& 568	& 672	& 375	& 1058	& 1158	& 408 \\
6.08\tablenotemark{a}	& 694	& 719	& 406	& 1228	& 1307	& 466 \\
6.14\tablenotemark{a}	& 696	& 766	& 391	& 1315	& 1401	& 525 \\
8.03\tablenotemark{a}	& 681	& 813	& 438	& 1298	& 1371	& 476 \\
8.04\tablenotemark{a}	& 617	& 642	& 375	& 1049	& 1151	& 397 \\
8.08\tablenotemark{a}	& 678	& 719	& 438	& 1178	& 1257	& 477 \\
8.11\tablenotemark{a}	& 754	& 781	& 406	& 1387	& 1476	& 575 \\
9.00\tablenotemark{a}	& 581	& 688	& 359	& 1142	& 1179	& 446 \\
10.00\tablenotemark{a}	& 661	& 719	& 391	& 1216	& 1258	& 472 \\
11.01\tablenotemark{a}	& 604	& 674	& 359	& 1091	& 1168	& 360 \\
11.09\tablenotemark{a}	& 754	& 751	& 375	& 1339	& 1428	& 573 \\
12.04\tablenotemark{a}	& 946	& 907	& 484	& 1927	& 2133	& 701 \\
13.13	& 994	& 969	& 531	& 2709	& 2566	& 1148 \\
15.01	& 2343	& 2205	& 750	& 3909	& 3079	& 1586 \\
16.03	& 1072	& 1032	& 563	& 2954	& 2830	& 1406 \\
17.02	& 1041	& 1001	& 594	& 2610	& 2425	& 711 \\
17.04	& 1056	& 954	& 563	& 2502	& 2440	& 984 \\
17.07	& 1087	& 1001	& 594	& 2556	& 2676	& 773 \\
19.06	& 994	& 954	& 500	& 2462	& 2345	& 734 \\
20.00	& 1041	& 1017	& 547	& 2787	& 2611	& 1305 \\
20.05	& 1154	& 1095	& 656	& 2574	& 2579	& 1281 \\
20.14	& 1058	& 1017	& 625	& 2541	& 2548	& 1289 \\
21.00	& 994	& 987	& 438	& 3107	& 3410	& 906 \\
21.03	& 979	& 907	& 453	& 2849	& 2814	& 828 \\
21.06	& 981	& 1016	& 500	& 2508	& 2612	& 742 \\
21.12	& 1043	& 969	& 469	& 4108	& 4175	& 1375 \\
22.04	& 992	& 1000	& 563	& 3776	& 3504	& 1422 \\
22.05	& 1092	& 985	& 438	& 4344	& 5049	& 1438 \\
23.07	& 886	& 876	& 438	& 1937	& 1909	& 492 \\
23.15	& 997	& 984	& 453	& 4114	& 5128	& 1094 \\
25.01	& 992	& 970	& 438	& 3016	& 3300	& 727 \\
25.04	& 944	& 939	& 438	& 3304	& 3096	& 648 \\
25.13	& 1025	& 985	& 469	& 4459	& 4315	& 1164 \\
25.15	& 1041	& 969	& 500	& 3463	& 3047	& 1242 \\
\enddata
\tablenotetext{a}{Final predicted crater measured algorithmically.}
\label{atab_crater}
\end{deluxetable}

\clearpage

\bibliography{Kumamoto_etal_2022_DART_R1}{}
\bibliographystyle{aasjournal}



\end{document}